\documentclass[aps,pra,twocolumn ,superscriptaddress]{revtex4-1}  
\usepackage{graphicx,epsfig}  
\usepackage{dcolumn}   
\usepackage{bm}        
\usepackage{amssymb}   
\usepackage{color}
\usepackage{mathtools}
\usepackage{float}
\usepackage{makecell}
\usepackage[section]{placeins}

\begin{document}

\title{Terahertz Generation through Photon Deceleration of Long-Wavelength Infrared Laser Pulses in Plasma}

\author{Srimanta Maity}
\email {srimantamaity96@gmail.com}
\affiliation{Atomic, Molecular and Optical Physics Division, Physical Research Laboratory, Navrangpura, Ahmedabad 380009, India}

\begin{abstract}

Efficient terahertz (THz) generation with high field amplitude and pulse energy is studied through the interaction of a single-color long-wavelength infrared (LWIR) laser pulse with gaseous targets. Particle-In-Cell (PIC) simulations are performed to investigate the underlying mechanism and analyze the properties of the emitted THz radiation. The results reveal that THz pulses at frequencies below 10 THz are generated via photon deceleration of the LWIR laser, driven by enhanced electron density accumulation at the pulse front in the self-modulated wakefield regime. The influence of key parameters, including target density, laser intensity, and propagation length, on nonlinear laser modulation and the resulting THz generation efficiency is analyzed. Important scalings governing the laser-to-THz energy conversion efficiency are identified from PIC simulations and validated through theoretical analysis. The study demonstrates a laser-to-THz energy conversion efficiency of approximately $4\%$, significantly exceeding previously reported values. For the laser parameters considered in this study, the energy of the emitted THz pulses is approximately 50 mJ. The findings of this study provide valuable insights for the development of next-generation high-energy THz sources.

\end{abstract}

\maketitle

\section{Introduction}\label{intro}

The development of tunable and intense terahertz (THz) radiation sources in the frequency range of $0.1$–$10$ THz has become an important area of research due to their wide range of applications \cite{tonouchi2007cutting}, including THz time-domain spectroscopy \cite{baxter2011terahertz, kampfrath2013resonant, berge2019terahertz}, probing explosives compounds \cite{chen2007absorption, davies2008terahertz}, medical imaging \cite{chan2007imaging, son2019potential}, security screening \cite{kemp2003security}, and atmospheric sensing \cite{clough2014toward}. Many of these applications require high-energy THz sources. Furthermore, high-energy THz pulses with ultrahigh electric field strengths ($>$ GV/m) provide access to nonlinear and relativistic light–matter interaction regimes that are difficult to achieve with conventional THz sources. However, despite significant progress in THz source development \cite{hafez2016intense}, the generation of high-energy and high-field THz pulses remains a major challenge.

Solid-state electronic devices, such as Schottky diodes \cite{crowe1992gaas} and quantum-cascade lasers \cite{faist1994quantum}, generally emit weak and very low-energy THz radiation. Subsequently, optically induced THz emitters emerged as reliable sources of THz radiation. For instance, optical rectification in nonlinear crystals \cite{loffler2005large} or tilted-pulse-front pumping \cite{hebling2004tunable} can produce low-energy THz radiation with a conversion efficiency of $\sim 10^{-2}$ and field strengths of $\sim 20$ MV/m. A significant enhancement in THz generation efficiency, with field strengths reaching $\sim 0.5$ GV/m, was later reported using a laser-driven large-size partitioned organic crystal \cite{vicario2014gv}. THz generation from lithium niobate crystals using femtosecond laser excitation has also been reported in recent studies \cite{lemery2020highly, wu2023generation}. However, in these methods, the laser-to-THz conversion efficiency at high pump intensities is limited by the self-phase modulation process. Additionally, these approaches are constrained by the damage threshold of the optical crystals and, in most cases, produce low-energy, narrow-bandwidth THz pulses with limited tunability.

Table-top laser–matter interactions in the relativistic-intensity regime offer a promising approach to developing high-energy, highly efficient THz sources \cite{hamster1993subpicosecond}. An overview of the current status and perspectives of laser-driven THz sources can be found in Refs. \cite{fulop2020laser, liao2023perspectives}. This approach provides several advantages over conventional technologies, including broadband THz generation, overcoming limitations related to material damage, enabling THz field strengths exceeding GV/m, and enhanced control over the THz generation process. For example, THz emission from thin foils irradiated by an intense laser pulse with an intensity of $\sim 10^{18}$ W/cm$^2$ has been reported \cite{herzer2018investigation}, producing THz pulse energies of $\sim 0.7$ mJ, field strengths of $\sim 10$ GV/m, and a bandwidth of $10$ THz. In relativistic laser–solid interactions, the main mechanisms for THz generation include linear mode conversion \cite{liao2015bursts} and transient currents at the target rear surface \cite{gopal2013observation}. The efficient generation of Terawatt-level, tunable THz radiation over the spectral range of 3–20 THz, with a laser-to-THz energy conversion efficiency of $\sim 1\%$, has been demonstrated via relativistic laser–foil interactions in recent experimental studies \cite{liao2020towards, ruan2026terawatt}. 

Gas plasma driven by two-color laser fields can also act as a THz emitter \cite{cook2000intense}, generating THz radiation via photoionization-induced radiation (PIR) \cite{kim2008coherent}. However, even at moderate laser intensities ($\sim 10^{14}$ W/cm$^2$), the laser-to-THz energy conversion efficiency of this method remains very low ($<10^{-3}$), resulting in $\mu$J-level THz pulse energies \cite{meng2016enhancement}. The wavelength dependence of THz generation from gas ionization driven by two-color lasers was also studied \cite{clerici2013wavelength}. In the relativistic interaction regime, high-field THz radiation with peak fields reaching the GV/cm level was theoretically predicted from the interaction of a single-color laser with gaseous plasma targets, where THz generation was attributed to the transverse residual momentum induced by local pump depletion \cite{chen2015high}. At higher intensities, a laser pulse interacting with a gas target can drive a strong plasma wave in its wake and accelerate electrons to relativistic energies via laser wakefield acceleration (LWFA) \cite{Tajima1979, RevModPhys.81.1229, maity2024parametric, maity2025coupling, maity2026laser}. These high-energy electrons, upon crossing the plasma–vacuum interface, emit high-energy radially polarized THz radiation via coherent transition radiation (CTR) \cite{leemans2003observation, leemans2004terahertz}. Through this mechanism, mJ-level THz pulse energies with energy conversion efficiencies exceeding $5\times10^{-3}$ have been reported \cite{dechard2018terahertz}. Mode conversion of laser wakefields in inhomogeneous plasmas driven by a femtosecond laser pulse can also generate THz radiation \cite{sheng2005emission}. A recent experimental study \cite{wang2024millijoule} reported that this mechanism can produce mJ-level THz radiation emitted in the backward direction, with a conversion efficiency of $\sim 2 \times 10^{-3}$. Theoretical studies \cite{nguyen2018broadband, PhysRevA.98.053415} also revealed that THz generation efficiency can be significantly enhanced by using mid- and far-infrared laser pulses interacting with gaseous plasma. The role of photoionization in THz generation from gaseous plasmas driven by ultraintense two-color lasers with near- to far-infrared carrier wavelengths was also studied using particle-in-cell (PIC) simulations \cite{dechard2019thz}. Amplification of a seeded THz wave in plasma via stimulated Raman scattering driven by a CO$_2$ pump laser was reported in a simulation study \cite{peng2023amplification}. Recently, a particle-in-cell (PIC) simulation study reported broadband THz generation with enhanced field strengths (hundreds of GV/m) from wakefields in near-critical-density plasmas driven by a single-color CO$_2$ laser pulse \cite{maity2025enhanced}. 

The present study investigates the efficient generation of THz radiation at frequencies below 10 THz through the interaction of long-wavelength infrared (LWIR) laser pulses with gaseous targets. Particle-In-Cell (PIC) simulations show that the generated THz radiation transmitted from the gaseous plasma reaches field strengths of hundreds of GV/m and pulse energies of $\sim 50$ mJ, corresponding to a laser-to-THz energy conversion efficiency of $\sim 4\%$. THz generation with such high field strengths, pulse energy, and conversion efficiency has not been reported in previous studies. In the present study, THz is generated via frequency downshifting (photon deceleration) of an infrared laser pulse interacting with the plasma electron density accumulation in the wakefield. Frequency shifting of laser pulses induced by plasma waves was first proposed theoretically in Ref. \cite{esarey1990frequency}. Later, experimental \cite{nie2018relativistic, nie2020photon} and theoretical \cite{zhu2022generation, lu2025generation} studies reported the generation of tunable infrared pulses through photon deceleration of pump lasers with wavelengths of $0.8-1.0~\mu$m in density-tailored plasmas. Building on these earlier works, the present study demonstrates that, by using an intense infrared laser pulse, such as a CO$_2$ laser \cite{polyanskiy2015chirped, polyanskiy2020demonstration, panagiotopoulos2020multi}, in a suitable parametric regime, a significant fraction of the pump laser energy can be efficiently frequency-downshifted to below 10 THz. The parametric dependence of the photon deceleration process for THz generation, particularly on plasma density, laser intensity, and plasma length, is investigated. The key scalings for maximizing the laser-to-THz energy conversion efficiency are identified through PIC simulations and verified by theoretical analysis.

The remainder of this paper is organized as follows. Section \ref{simu} describes the PIC simulation setup and the key parameters. Section \ref{rd} presents the simulation results and analysis. Section \ref{modulation_section} highlights the characteristics of self-modulation of the laser envelope as it propagates through the gas target, Section \ref{THz_section} presents a detailed analysis of photon-deceleration-driven THz generation, and Section \ref{3D_section} discusses the three-dimensional effects on the observed physical processes. Finally, Section \ref{summary} summarizes the main results with concluding remarks.

\section{Particle-In-Cell Simulation Setup}\label{simu}

Particle-In-Cell (PIC) simulations were performed using the open-source, fully relativistic, and massively parallelized PIC code, EPOCH \cite{arber2015contemporary, bennett2017users}. In these simulations, the second-order Yee scheme \cite{yee1966numerical} was employed as the field solver, while the Boris rotation algorithm \cite{boris1970relativistic}, based on a modified leapfrog method, was used as the particle pusher. The simulations were carried out in a two-dimensional (2D) rectangular Cartesian ($x$–$y$) geometry, with the simulation box extending from $0$ to $150\lambda_0$ along $\hat{x}$ and from $-50\lambda_0$ to $50\lambda_0$ along $\hat{y}$. Here, $\lambda_0 = 10.6~\mu$m represents the laser wavelength considered in this study. The grid size was chosen as $\Delta x = \lambda_0/40 = 0.265~\mu$m along $\hat{x}$ and $\Delta y = \lambda_0/30 = 0.353~\mu$m along $\hat{y}$. The simulation time step, $\Delta t$, was set in accordance with the CFL (Courant-Friedrichs-Lewy) condition \cite{courant1967partial}, $\Delta t \leq C/\left(c\sqrt{1/\Delta x^2 + 1/\Delta y^2}\right)$, where $C$ is the Courant number and $c$ denotes the speed of light in vacuum. In the simulations, $C$ was chosen to be $0.99$, i.e., close to unity, in order to minimize numerical dispersion, resulting in a simulation time step of $\Delta t = 0.7$ fs. Sixteen simulation particles (macro-particles) were initialized per cell. Open boundary conditions were applied for both particles and electromagnetic (EM) waves, allowing them to freely propagate through the boundaries without reflection.

A long-wavelength infrared (LWIR) laser pulse with a wavelength of $\lambda_0 = 10.6~\mu$m was launched from the left boundary of the simulation box ($x = 0$) and was set to propagate along $\hat{x}$, with its electric field polarized along $\hat{z}$. The incident laser frequency was $\nu_0 \simeq 28.28$ THz, and the corresponding critical density was $n_c \simeq 9.9 \times 10^{18}$ cm$^{-3}$. The laser had Gaussian profiles in both the transverse and longitudinal directions, with a beam waist of $w_{0} = 30~\mu$m and a full width at half maximum (FWHM) pulse duration of $\tau_{\rm fwhm} = 0.3$ ps. The laser was initially focused at $x = 200~\mu$m, i.e., at the beginning of the target plateau. In the present simulations, the peak value of the normalized vector potential at the vacuum focus was taken as $a_0 = 8.5493 \times 10^{-10}\sqrt{I_0}$ [W/cm$^2$] $\lambda_0$ [$\mu$m] $= 5.0$, corresponding to a peak intensity of $I_0 = 3.044 \times 10^{17}$ W/cm$^2$ and an on-target laser energy of $1.3$ J. However, $a_0$ was also varied over the range from $1.0$ to $7.5$ to study the effect of the initial laser intensity (or energy).

Helium (He) gas with a trapezoidal density profile along $\hat{x}$ and no variation along $\hat{y}$ was considered as the target. The longitudinal density profile consisted of a $10\lambda_0$ upramp, followed by a plateau length of $75\lambda_0$, and a downramp length of $10\lambda_0$. The target plateau started at $x = 200~\mu$m, maintaining a $\sim 100~\mu$m vacuum region between the left boundary of the simulation box and the beginning of the gas target. The lengths of the upramp and downramp were chosen to provide smooth transitions between the vacuum and the target plateau. They do not affect the physical process discussed in this work, which is primarily governed by the laser parameters and the plasma plateau parameters, such as the plasma density and plateau length. The field-ionization module was employed in the PIC simulations, incorporating both multiphoton \cite{delone2000multiphoton} and tunneling ionization \cite{ammosov1986tunnel, krainov1995theory} processes to model the laser-driven ionization of neutral He atoms. Although the target was initialized as neutral helium gas with density $n_{\rm He}$, for convenience, the corresponding fully ionized electron density, $n_0 = 2n_{\rm He}$, is used throughout this paper to express the initial target density. In the simulations, $n_0$ was varied from $0.01n_c$ to $0.1n_c$.

The laser parameters considered in this study are mainly relevant to TW-class CO$_2$ laser systems \cite{polyanskiy2015chirped, polyanskiy2020demonstration, panagiotopoulos2020multi}. The gas target parameters used in the simulations can be obtained experimentally using a gas-jet setup or a capillary gas-cell configuration. Similar laser and plasma conditions have been used in experiments on laser wakefield acceleration of electrons \cite{zgadzaj2024plasma}.


\section{Results and Discussion}
\label{rd}

\subsection{Nonlinear Modulation of Laser Pulse}
\label{modulation_section}

\begin{figure*}
  \centering
  \includegraphics[width=0.84\textwidth]{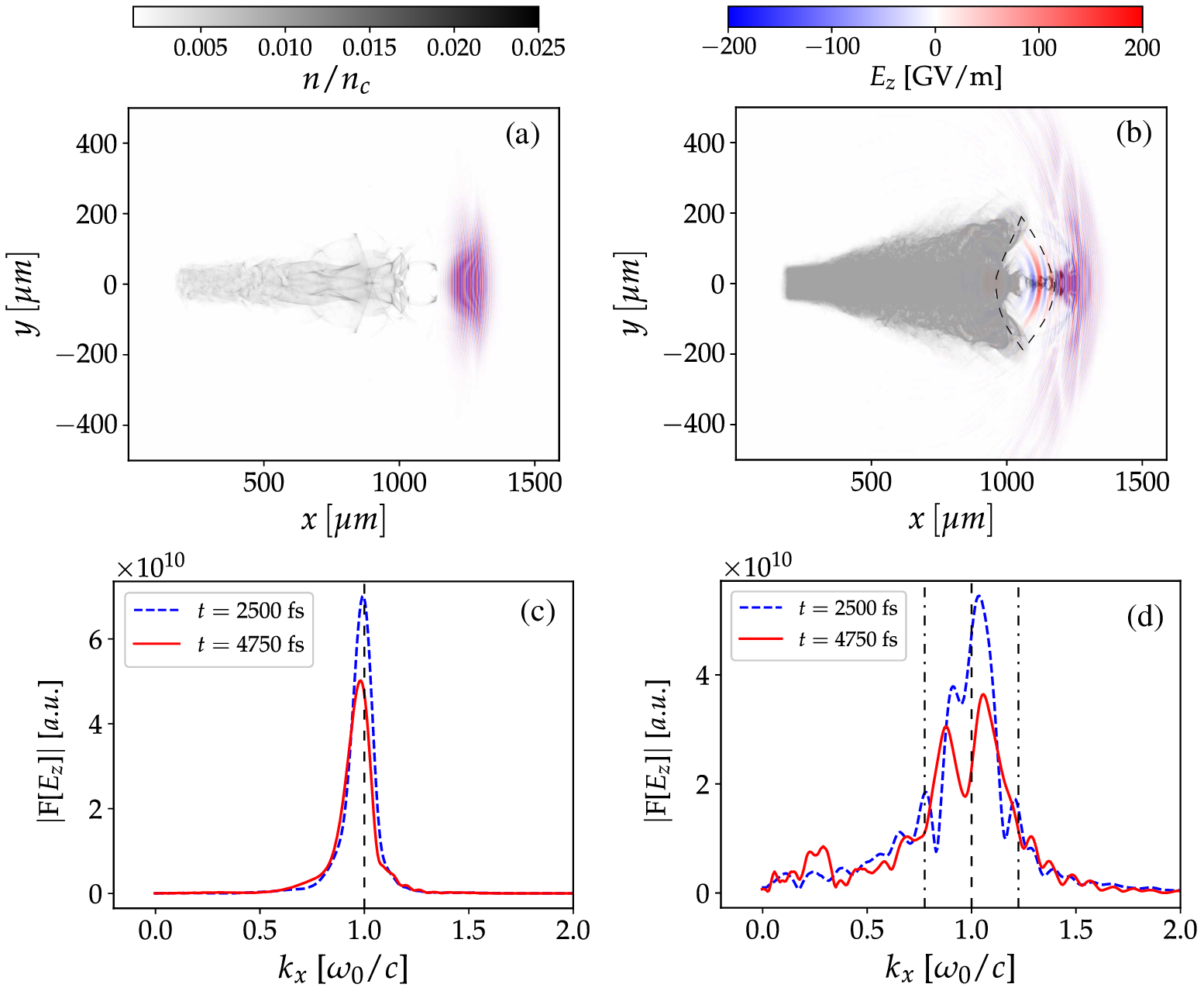}
  \caption{Electron density $n$ (gray colormap) and $z$-polarized electric field $E_z$ (blue--white--red colormap) at a simulation time $t = 4750$ fs (corresponding to $\sim 200~\mu$m of propagation after exiting the plasma channel) for two initial plasma densities: (a) $n_0 = 0.01n_c$ and (b) $n_0 = 0.05n_c$. The peak value of the initial normalized vector potential was fixed at $a_0 = 5.0$ in both cases. The black dashed contour in (b) provides a visual guide for the generated longer-wavelength pulses trailing the main transmitted laser pulse. (c) and (d) show the Fourier spectra of on-axis z-polarized electric field $E_z(x)$ at $t = 2500$ fs and $4750$ fs for cases (a) and (b), respectively. The black dashed vertical lines in (c) and (d) mark the laser wave vector ($\omega_0/c$) in vacuum, while the black dash-dotted lines in (d) indicate the Stokes and anti-Stokes sidebands.}
\label{fig_space_fft}
\end{figure*}


\begin{figure}
  \includegraphics[width=3.0in]{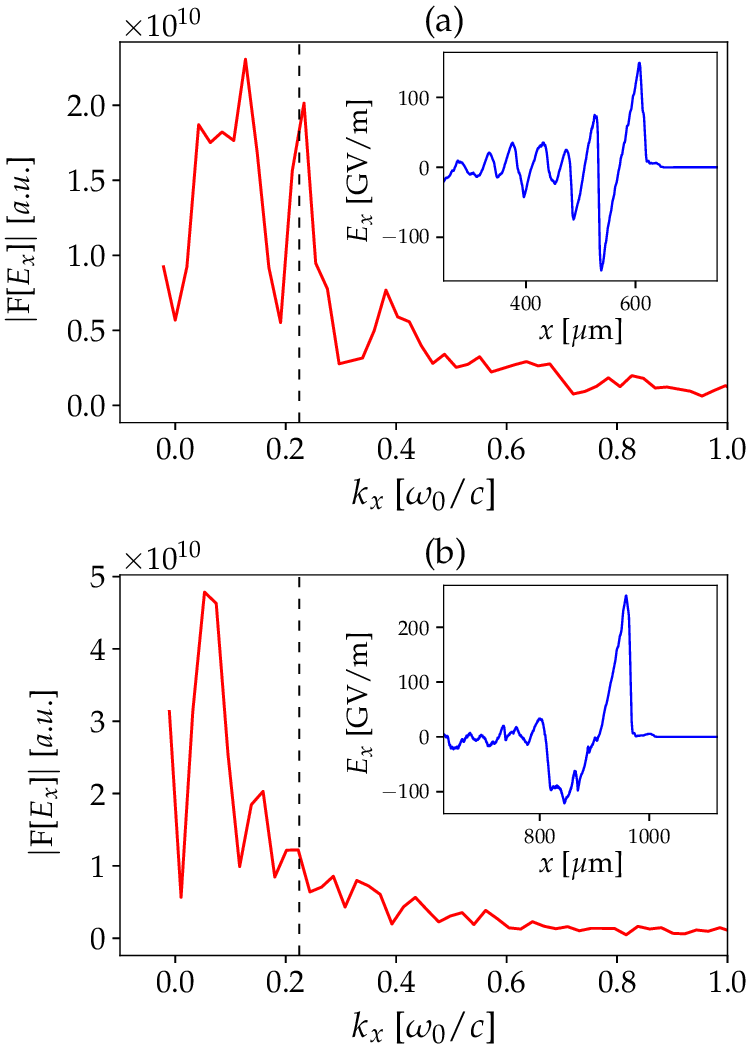}
  \caption{Fourier spectra of the on-axis longitudinal electric field $E_x$ for a particular case with $n_0 = 0.05n_c$ and $a_0 = 5.0$ at times (a) $t = 2500$ fs and (b) $t = 3750$ fs. The insets show the corresponding $E_x$ as a function of $x$.}
\label{fig_Ex_fft}
\end{figure}


\begin{figure*}
  \centering
  \includegraphics[width=0.84\textwidth]{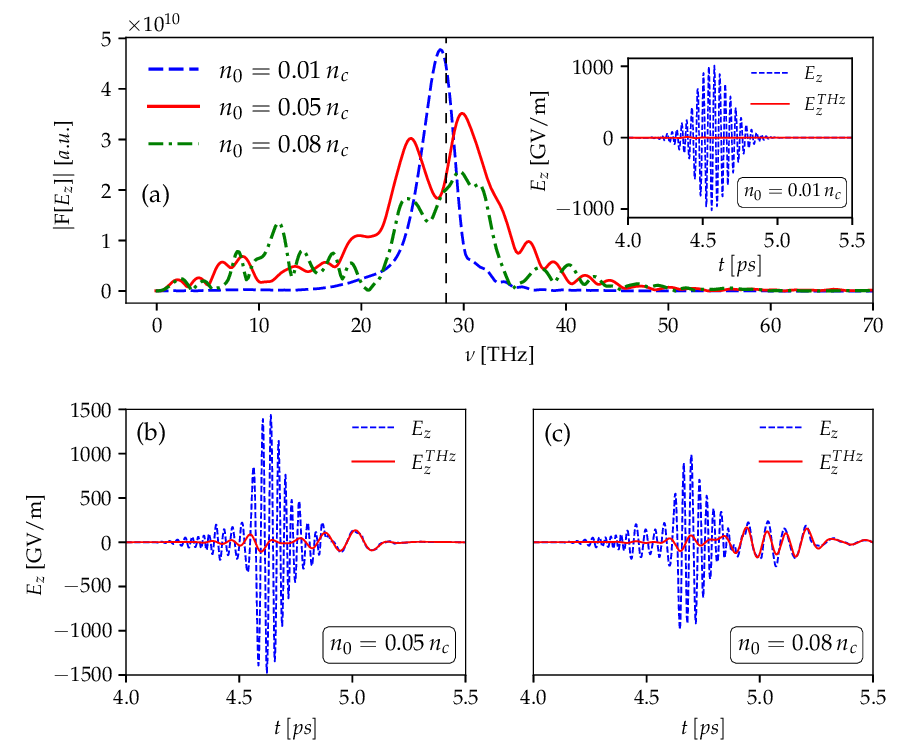}
  \caption{(a) Fourier spectra of the on-axis $z$-polarized electric field $E_z(t)$ at $x = 1200~\mu$m, i.e., $100~\mu$m beyond the plasma channel, for three values of the initial plasma density $n_0$. The inset of (a) and (b)–(c) depict the on-axis field $E_z(t)$ (blue dashed line) and the THz field $E_z^{THz}$ (red solid line), extracted from $E_z(t)$ by frequency filtering in the range $\nu < \nu_0/3$, for the corresponding cases.}
\label{fig_time_fft}
\end{figure*}


As an intense long-wavelength infrared (LWIR) laser pulse propagates through the He gas target, its leading edge, which has sufficient intensity to fully ionize the helium atoms, creates a plasma channel along its propagation path. The interaction of the laser pulse with the generated plasma gives rise to several nonlinear phenomena \cite{sprangle1990nonlinear, bulanov1995two}, including nonlinear wakefield (bubble) \cite{pukhov2002laser} excitation and self-phase modulation of the laser pulse \cite{joshi1981forward}. For a sufficiently high plasma density satisfying the condition $\lambda_{p0} < c\tau_{\rm fwhm}$, the laser pulse undergoes self-modulation instability \cite{andreev1992resonant, esarey1994envelope}, leading to the generation of Stokes ($\omega_0 - \omega_{p0}$ ; $k_0 - k_{p0}$) and anti-Stokes ($\omega_0 + \omega_{p0}$ ; $k_0 + k_{p0}$) sidebands in the transmitted laser spectrum \cite{antonsen1992self, zgadzaj2024plasma}. Here, $k_0=\omega_0/c$ is the wavenumber of the incident laser pulse, $\omega_{p0}=\sqrt{n_0 e^2/(\epsilon_0 m)}$ is the ambient plasma frequency, and $k_{p0}=\omega_{p0}/c$ is the wavenumber associated with the plasma wave, where $\epsilon_0$ denotes the permittivity of free space and $m$ is the electron mass. A detailed analysis of the evolution of these nonlinear processes is presented below. 

 Snapshots of the electron density $n$ (gray colormap) and the $z$-polarized electric field $E_z$ (blue--white--red colormap) at a particular simulation time after the laser exits the target are shown in Fig. \ref{fig_space_fft}(a)-(b) for two different initial plasma densities, $n_0 = 0.01n_c$ and $0.05n_c$, respectively. In both cases, plasma channels are formed along the laser propagation path. In addition, the transmitted laser field ($E_z$) is modified significantly for the case with $n_0 = 0.05n_c$ compared to the case with $n_0 = 0.01n_c$. For the former case, the plasma wavelength is $\lambda_{p0} \approx 47~\mu$m, which is smaller than the FWHM laser spot size, $w_{\rm fwhm} \sim 50~\mu$m. As a result, the laser pulse undergoes transverse modulation and broadens. The laser pulse also gets modulated in the longitudinal direction for $n_0 = 0.05n_c$, as in this case $\lambda_{p0}$ is smaller than the pulse length $c\tau_{\rm fwhm} \approx 90~\mu$m. Furthermore, for $n_0 = 0.05n_c$, a few-cycle pulse with a longer wavelength, indicated by the black dashed contour in Fig. \ref{fig_space_fft}(b), is observed behind the main transmitted laser field. Such a signal is absent in the case with $n_0 = 0.01n_c$.

 To understand the evolution of the laser electric field, the Fourier transform of the on-axis $E_z$ is computed for two initial target densities at two instants of time: $t = 2500$ fs, when the laser is inside the plasma, and $ t = 4750$ fs, after it exits the target. The results are shown in Fig. \ref{fig_space_fft}(c)-(d). It is found that for $n_0 = 0.01n_c$, the wavenumber of the laser field ($E_z$) remains almost unchanged during propagation, as shown in Fig. \ref{fig_space_fft}(c). This is because for this case $c\tau_{\rm fwhm} < \lambda_{p0}$, and thus the laser pulse does not satisfy the condition for self-modulation instability.

However, for the case with higher initial target density, $n_0 = 0.05n_c$, the wavenumber spectrum of $E_z$ changes significantly during the propagation, as shown in Fig. \ref{fig_space_fft}(d). At $t = 2500$ fs, corresponding to a propagation distance of about $\sim 600~\mu$m, along with the main peak at $k_0 = \omega_0/c$, two extra sidebands appear at $k_0-k_{p0}$ and $k_0+k_{p0}$. These are the usual Stokes and anti-Stokes Raman sidebands, respectively. However, at a later time, $t = 4750$ fs, these Raman sidebands, along with the main peak at $k_0$, disappear in the transmitted laser spectrum. Instead, two strong peaks appear around the incident laser wavenumber $k_0$. In addition, a peak below $k_x \sim 0.3k_0$, corresponding to the $<10$ THz region, emerges at this later stage. The properties and the origin of this THz field are discussed in Sec. \ref{THz_section}.

To analyze the origin of the change in the Stokes and anti-Stokes sidebands for the higher density case, the on-axis longitudinal electric field ($E_x$) and its Fourier spectrum are studied for $n_0 = 0.05n_c$ at two different times, $t = 2500$ fs and $3750$ fs. The latter corresponds to the point where the laser reaches the end of the plasma plateau. The results are shown in Fig. \ref{fig_Ex_fft}(a)-(b). It is seen that up to a propagation distance of about $\sim 600~\mu$m, an almost sinusoidal wakefield with many cycles is excited. The amplitude and wavelength are larger for the cycle formed just behind the laser pulse, i.e., the first bubble, as shown in the inset of Fig. \ref{fig_Ex_fft}(a). Consequently, two strong peaks appear in the Fourier spectrum. One peak occurs at $k\lesssim k_{p0}$, arising from the slightly larger-wavelength first bubble behind the laser, while the other is located at the wavenumber of the linear plasma wave, $k_{p0} = 2\pi/\lambda_{p0}$. Therefore, up to this stage, the laser pulse is primarily modulated by the plasma wave at the linear wavelength, and the usual Raman sidebands appear in the Fourier spectrum of $E_z$, as shown by the blue dashed line in Fig. \ref{fig_space_fft}(d).

At a later time, as the laser propagates deeper into the target, the inset of Fig. \ref{fig_Ex_fft}(b) shows that only one cycle of $E_x$, corresponding to the first bubble, remains, with a much larger amplitude and wavelength. This occurs because the wakefield enters a strongly nonlinear, turbulent regime as the laser peak intensity increases due to relativistic self-focusing during propagation \cite{antonsen1992self, sprangle1992propagation}. As a result, only one dominant peak appears in the Fourier spectrum at $k_x \approx 0.05 k_0$, corresponding to the wavenumber $k_{pn}$ of the first bubble. Therefore, at this later stage of evolution, the usual Raman sidebands disappear, and the laser envelope becomes nonlinearly modulated. Consequently, the transmitted laser field exhibits modified Raman sidebands at wavenumbers $k_0 \pm k_{pn}$, as illustrated by the red solid line in Fig. \ref{fig_space_fft}(d).

\subsection{THz Generation}
\label{THz_section}
 
In this section, the longer wavelength pulse observed behind the transmitted laser pulse, with wavenumber $<0.3k_0$ and corresponding frequency $<10$ THz, as highlighted in Fig. \ref{fig_space_fft}(b) and (d), is analyzed in detail. To study the temporal profile of the generated THz signal, the Fourier transform of the on-axis $E_z(t)$ at $x = 1200~\mu$m, i.e., $100~\mu$m away from the target, is performed. The THz signal is then extracted by frequency filtering in the range $\nu < \nu_0/3$, where $\nu~(\nu_0)=\omega~(\omega_0)/2\pi$. The results for different initial target densities are presented in Fig. \ref{fig_time_fft}.

The frequency spectra shown in Fig. \ref{fig_time_fft}(a) reveal a strong dependence on the initial target density. For the lower density case, $n_0 = 0.01n_c$, only a single peak appears at the incident laser frequency, $\nu_0 \sim 28.3$ THz. However, as the initial plasma density increases, two dominant peaks appear around the incident laser frequency, which correspond to the nonlinear modulation of the laser envelope. These results are consistent with the wavenumber spectra shown in Fig. \ref{fig_space_fft}(c)-(d). In addition, peaks also appear in the low-frequency THz region ($\nu \lesssim 10$ THz), and the power in this region increases with increasing target density. This indicates that a significant fraction of the laser power is converted into low-frequency THz radiation.

The extracted THz signal, $E_z^{\mathrm{THz}}$ (within the frequency range $\nu < \nu_0/3$), together with the net transmitted field $E_z(t)$ at $x = 1200~\mu$m, is shown for the three initial plasma densities in the inset of Fig. \ref{fig_time_fft}(a) and Figs. \ref{fig_time_fft}(b)-(c). It is observed that no THz radiation is generated for the case with $n_0 = 0.01n_c$, whereas strong THz emission with a field amplitude of about $\sim 100$ GV/m is observed for both $n_0 = 0.05n_c$ and $0.08n_c$. In both cases, the generated THz signal is delayed with respect to the main transmitted pulse. This agrees with the observation highlighted in Fig. \ref{fig_space_fft}(b), where the longer-wavelength THz pulse trails the main pulse. It is also seen that the number of oscillation cycles in the extracted THz waveform increases as the initial plasma density $n_0$ increases.


\begin{figure*}
  \centering
  \includegraphics[width=0.84\textwidth]{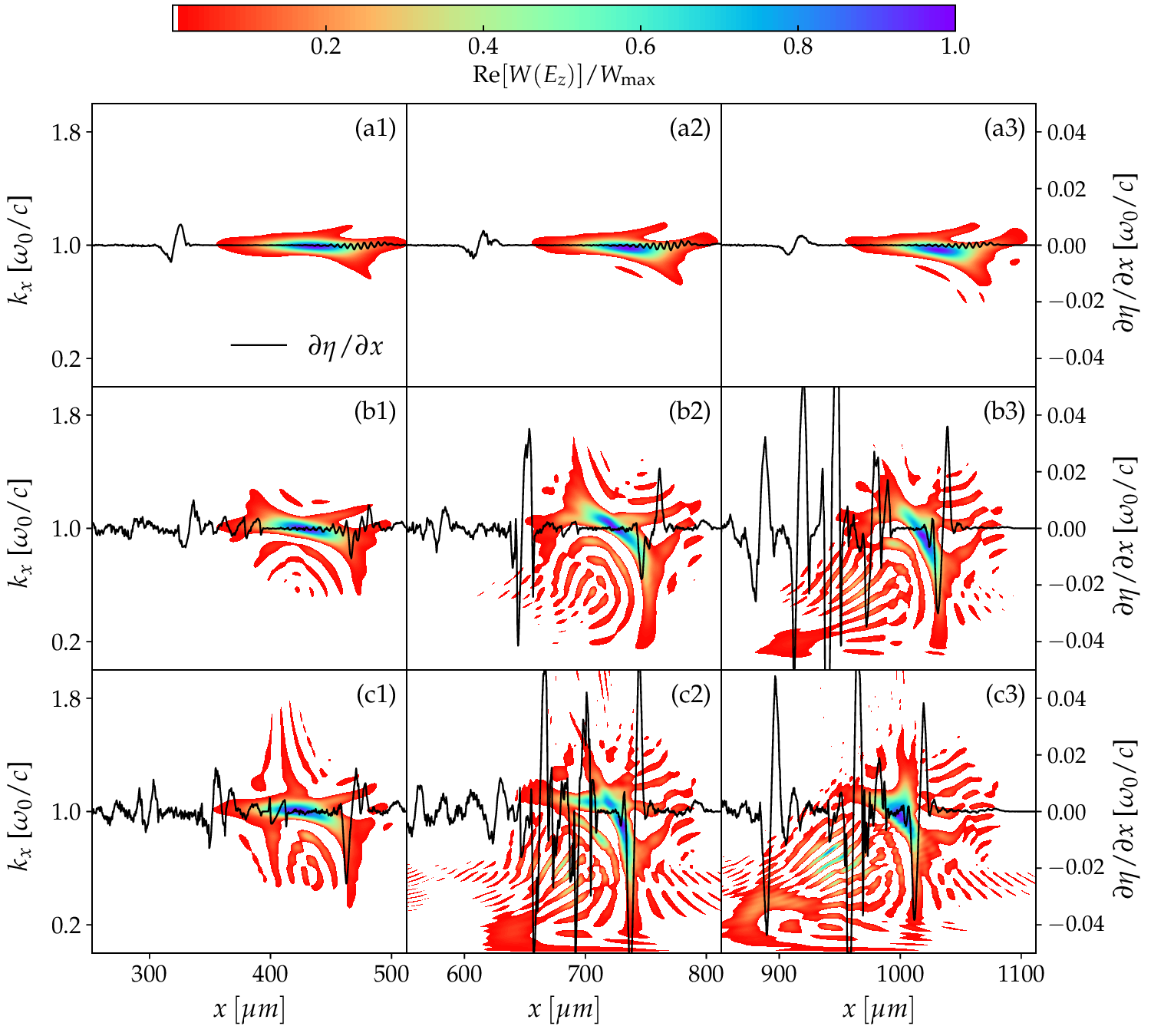}
  \caption{Wigner spectrum (colorbar) of the on-axis electric field $E_z$, defined as
$W(E_z) = \int_{-\infty}^{+\infty} E_z(x+x'/2)E_z^*(x-x'/2)e^{-ikx'}dx'$, along with the gradient of the on-axis refractive index, $\partial \eta/\partial x$ (black solid line). The refractive index is defined as $\eta=\sqrt{1-n/(n_c\gamma)}$, where $n/n_c$ and $\gamma=\sqrt{1+\langle a^2\rangle}$ are obtained from the PIC simulation data. (a1)–(a3) correspond to an initial plasma density $n_0=0.01n_c$ at simulation times $t=2000$, $3000$, and $4000$ fs, respectively. (b1)–(b3) and (c1)–(c3) show the corresponding results for $n_0=0.05n_c$ and $0.08n_c$, respectively.}
\label{fig_wigner}
\end{figure*}


\begin{figure}
  \centering
  \includegraphics[width=3.0in]{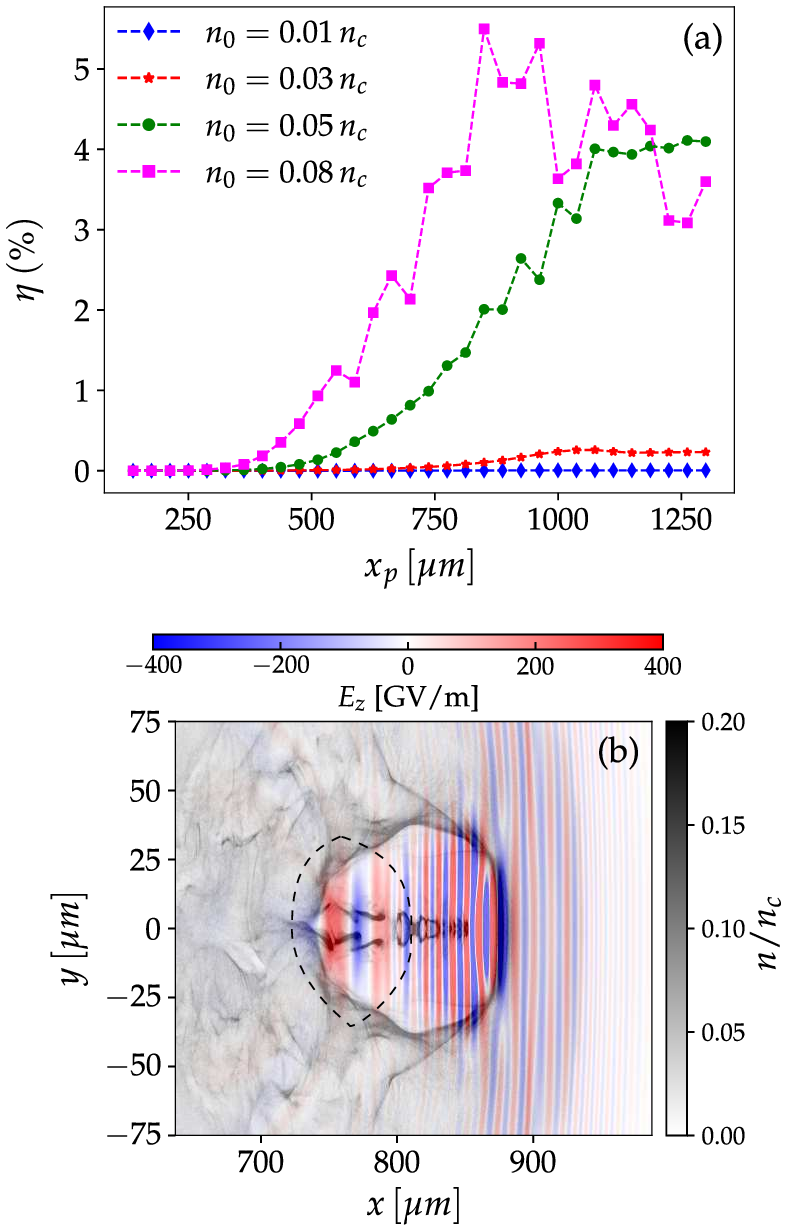}
  \caption{(a) Laser-to-THz energy conversion efficiency, $\eta = \int |E_z^{\rm THz}|^2 dxdy / \int |E_z^{\rm laser}|^2 dxdy$, as a function of laser propagation distance $x_p = ct$ for various initial plasma densities $n_0$ at a fixed $a_0 = 5.0$. (b) Electron density $n$ (gray colormap) and electric field $E_z$ (blue–white–red colormap) at $t = 3500$ fs for the case with $n_0 = 0.08n_c$. The black dashed contour marks the region where the redshifted signals (THz radiation), lagging behind the main laser pulse, reach the electron sheath at the rear of the bubble.}
\label{fig_eta_n0}
\end{figure}


\begin{figure}
  \includegraphics[width=3.0in]{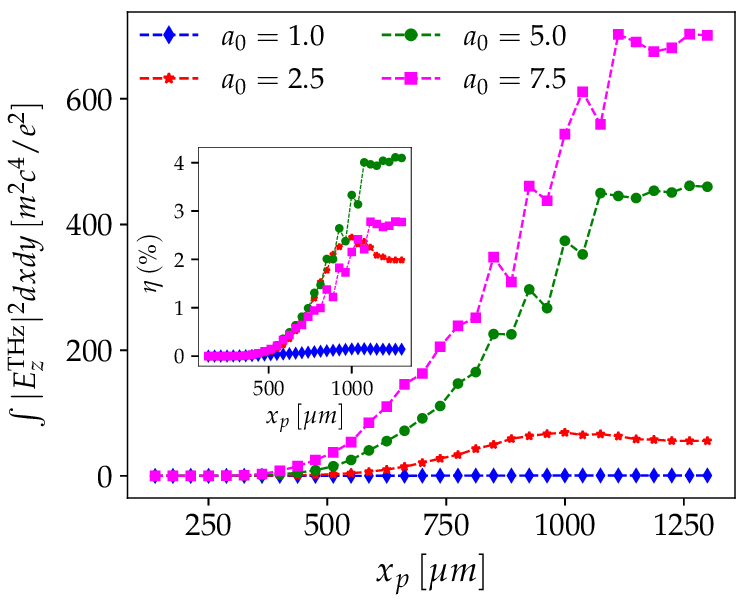}
  \caption{THz energy as a function of propagation distance $x_p = ct$ for various initial laser intensities ($a_0$) at a fixed initial plasma density $n_0 = 0.05n_c$. The inset shows the corresponding laser-to-THz energy conversion efficiency ($\eta$) as a function of propagation distance $x_p$. }
\label{fig_eta_a0}
\end{figure}
\begin{figure}
  \centering
  \includegraphics[width=3.0in]{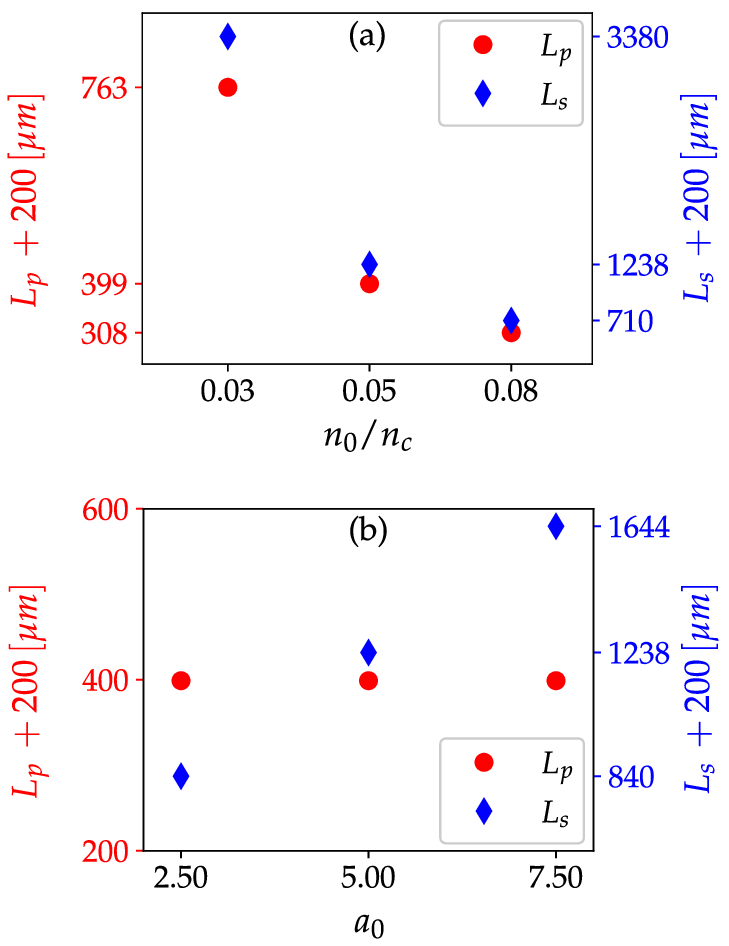}
  \caption{Minimum propagation distance $L_p$ required for laser photons to redshift below the frequency threshold $\nu < \nu_0/3$, and propagation distance $L_s$ required for the generated redshifted (THz) signal to lag behind the original laser pulse due to group velocity dispersion (GVD) and reach the electron sheath at the rear of the bubble. The theoretically calculated values of these two parameters are shown in (a) as a function of initial plasma density $n_0$ for a fixed $a_0 = 5.0$, and in (b) as a function of $a_0$ for a fixed $n_0 = 0.05n_c$.}
\label{fig_theory}
\end{figure}


The THz generation observed in this study is driven by the photon deceleration of a laser pulse propagating through plasma in the self-modulated wakefield (SMWF) regime. When a driver laser pulse satisfying $c\tau_{\rm fwhm} > \lambda_{p0}$ propagates through plasma, it resonantly excites a nonlinear wakefield via the self-modulation instability \cite{andreev1992resonant}. This wakefield consists of spatiotemporally varying regions of enhanced electron density accumulation and depletion. Consequently, the local plasma refractive index $\eta(x) = \sqrt{1-n(x)/(n_c\gamma)}$ \cite{sprangle1990nonlinear} is strongly modulated, leading to a local frequency shift of the driver laser pulse. Here, $\gamma=\sqrt{1+\langle a^2\rangle}$ represents the relativistic Lorentz factor associated with the electron dynamics in the laser field, where $a = eE_z/mc\omega_0$ denotes the normalized vector potential of the laser field. The change in the laser frequency and wavenumber in the presence of a plasma wave can be expressed as \cite{esarey1990frequency} 

\begin{equation}
    \Delta \omega = -v_g\tau \left(\frac{\omega_{p0}^2}{2\omega_0}\right) \frac{1}{n_0} \frac{\mathrm{d}}{\mathrm{d}\zeta}\delta n~~,
\label{eq1}    
\end{equation}

\begin{equation}
    \Delta k = -\tau \left(\frac{\omega_{p0}^2}{2\omega_0}\right) \frac{1}{n_0} \frac{\mathrm{d}}{\mathrm{d}\zeta}\delta n~~,
\label{eq2}
\end{equation}

\noindent
where $\Delta \omega = \omega - \omega_0$ and $\delta k = k - k_0$, while $\delta n = n - n_0$ denotes the plasma density perturbation. Here, $\zeta = x - v_g t$ and $\tau = t$ are the transformed coordinates, where $v_g$ represents the group velocity of the laser pulse propagating through the plasma medium. In deriving Eqs.~(\ref{eq1}) and (\ref{eq2}), it is assumed that the phase velocity ($v_p$) of the plasma wave is equal to the group velocity ($v_g$) of the laser pulse. Equations~(\ref{eq1}) and (\ref{eq2}) can be rewritten in terms of the plasma refractive index ($\eta$) as

\begin{equation}
\Delta \omega = v_g\tau \left(\gamma \frac{\omega_{p0}^2}{\omega_0} \frac{n_c}{n_0}\right)\eta \frac{\partial \eta}{\partial\zeta}~~,
\label{eq3}
\end{equation}

\begin{equation}
\Delta k = \tau \left(\gamma \frac{\omega_{p0}^2}{\omega_0} \frac{n_c}{n_0}\right)\eta \frac{\partial \eta}{\partial\zeta}~~,
\label{eq4}
\end{equation}

\noindent
where $n_c$ is the critical plasma density corresponding to the laser frequency $\omega_0$. From Eqs.~(\ref{eq3}) and (\ref{eq4}), it can be seen that the laser frequency and wavenumber will be downshifted (photon deceleration) in regions where $\partial \eta/\partial x < 0$ and upshifted (photon acceleration) in regions where $\partial \eta/\partial x > 0$.

During the highly nonlinear stage of the wakefield, commonly referred to as the bubble regime \cite{pukhov2002laser}, nearly all plasma electrons are expelled from the wake, forming an ion cavity surrounded by a dense electron sheath resulting from nonlinear electron accumulation. Consequently, the refractive index remains nearly uniform ($\partial \eta/\partial x \approx 0$) inside the bubble and exhibits significant variation only at its boundaries. Since the driver laser pulse interacts primarily with the electron-density accumulation at its front, where $\partial \eta/\partial x < 0$, it undergoes enhanced frequency and wavenumber downshifting (photon deceleration). The remaining part of the laser pulse mainly resides within the bubble, where $\partial \eta/\partial x \approx 0$. Therefore, the overall spectral evolution of the driver laser pulse is governed predominantly by photon deceleration rather than photon acceleration. These features are revealed in the PIC simulations, as shown in Fig.~\ref{fig_wigner}.

To analyze the local change in the laser wavenumber, the Wigner transformation of the $z$-polarized electric field, $E_z$, defined as $W(E_z) = \int_{-\infty}^{+\infty} E_z(x+x'/2)E_z^*(x-x'/2)e^{-ikx'}dx'$, has been performed at different instants of time during laser propagation in the plasma. The resulting Wigner distributions are presented in Fig.~\ref{fig_wigner} for different initial plasma densities, $n_0$, with a fixed $a_0 = 5.0$. The refractive index gradient ($\partial \eta/\partial x$) for the corresponding cases is also shown. 

For the case with $n_0 = 0.01n_c$, for which $c\tau_{\rm fwhm}<\lambda_{p0}$, the electron-density accumulation at the front of the laser pulse is negligible, resulting in $\partial \eta/\partial x \approx 0$ over the entire pulse length throughout the propagation. Consequently, no significant change in the laser wavenumber is observed in this case, as shown in Fig.~\ref{fig_wigner}(a1)–(a3). 

As the initial plasma density is increased to $n_0 = 0.05n_c$, for which $c\tau_{\rm fwhm}>\lambda_{p0}$, the wakefield amplitude as well as the electron-density accumulation at the front of the laser pulse are significantly enhanced by the self-modulation instability. This causes $\partial \eta/\partial x$ to become sufficiently large to drive the wavenumber (and frequency) shift of the laser pulse, as shown in Fig.~\ref{fig_wigner}(b1)–(b3). It is observed that the front and central high-intensity regions of the laser pulse encounter a negative refractive index gradient, leading to a dominant downshift in the laser wavenumber. In contrast, the tail region of the laser pulse exhibits both downshifting and upshifting in wavenumber as it propagates through regions of positive and negative refractive index gradients generated by electrons self-injected into the bubble. As the density gradient (and consequently $\partial \eta/\partial x$) continues to increase throughout propagation due to the self-modulation instability, the wavenumber at the front of the laser pulse keeps decreasing with propagation distance. Furthermore, the group velocity of an electromagnetic wave in plasma depends on its frequency as $v_g(\omega)=c\sqrt{1-\omega_{p0}^2/\omega^2}$, i.e., the group velocity decreases with decreasing frequency. Thus, the downshifted radiation produced near the front of the laser pulse propagates with a lower group velocity, causing it to lag behind the main pulse and eventually separate from it, as depicted in Fig.~\ref{fig_wigner}(b3). 

A similar feature is also observed for $n_0 = 0.08n_c$, as shown in Fig.~\ref{fig_wigner}(c1)–(c3). However, in this case, owing to the higher initial plasma density, the wakefield amplitude and consequently the electron density accumulation grow much faster. As a result, the rate of downshifting of the laser wavenumber increases significantly. Consequently, due to negative group velocity dispersion (GVD), the downshifted radiation separates from the main laser pulse at an earlier stage of propagation, as evident in Fig.~\ref{fig_wigner}(c2). At a later stage, the average wavenumber of the downshifted radiation is observed to increase again, as shown in Fig.~\ref{fig_wigner}(c3). This occurs because the downshifted radiation reaches the rear of the bubble, where it encounters a positive refractive index gradient arising from the negative density gradient of the electron sheath. 

The THz generation mechanism in the present study distinctly differs from other THz generation mechanisms, such as CTR \cite{leemans2003observation, dechard2019thz} and coherent forward radiation from laser-accelerated plasma electrons \cite{pak2023multi}. Unlike CTR, which is generated when the LWFA-accelerated electrons cross the plasma-vacuum boundary, the THz field in the present study develops continuously inside the plasma. The coherent electron-driven radiation mechanism reported in Ref. \cite{pak2023multi} originates from the sudden acceleration and scattering of low-energy plasma electrons at the rear of the plasma bubble, producing far-field, radially polarized THz radiation away from the laser pulse and its propagation axis. In contrast, in the present study, the generated THz field is aligned with the laser propagation axis, has the same polarization, and initially remains attached to the laser pulse. At later times, it separates from the laser due to negative group-velocity dispersion. As a result, the THz field observed in the present study has its highest amplitude on the laser propagation axis, whereas the forward THz radiation from the electron-driven mechanisms vanishes on axis. These features, together with the continuous downshifting of the laser frequency (and wavenumber) into the THz range, as shown in Fig. \ref{fig_wigner}, support photon deceleration as the primary THz generation mechanism in the present study.

The generated THz radiation has a broad spectrum extending below $10$ THz, resulting from the continuous downshift of the laser frequency (and wavenumber) during propagation through the target. The THz spectrum can be controlled by varying the initial plasma density and the plasma plateau length. The plasma density affects the evolution of the wakefield and the resulting refractive-index gradient, which governs the rate of photon deceleration. The plasma plateau length, in turn, determines how long this process continues and how far the laser frequency shifts toward the THz range. These features are also evident in the Wigner spectra in Fig. \ref{fig_wigner}.

Laser-to-THz energy conversion efficiency, defined as $\eta = \int |E_z^{\rm THz}|^2 dxdy / \int |E_z^{\rm laser}|^2 dxdy$, has been calculated over the laser propagation distance $x_p$ from the PIC simulation data. Here, $E_z^{\rm laser}$ is the initial laser electric field. The THz field $E_z^{\rm THz}(x,y)$ is obtained by Fourier filtering $E_z(x,y)$ in the range $k < k_0/3$, corresponding to frequencies below $9.5$ THz. The conversion efficiency $\eta$ is shown in Fig.~\ref{fig_eta_n0}(a) for various initial plasma densities with a fixed $a_0 = 5.0$.

As discussed earlier, for $n_0 = 0.01n_c$, no THz radiation is generated throughout the propagation distance. For $n_0 \geq 0.03n_c$, $\eta$ increases with increasing initial plasma density. For intermediate plasma densities, such as $n_0 = 0.03n_c$ and $0.05n_c$, the conversion efficiency increases monotonically throughout the entire plasma plateau. In contrast, for a higher initial plasma density, e.g., $n_0 = 0.08n_c$, $\eta$ reaches its maximum value ($\sim 5.2\%$) at $x_p \approx 750~\mu$m, well before the end of the plasma plateau. This behavior arises because the laser-to-THz conversion process is initiated much earlier in this case. Consequently, the generated THz radiation lags behind the main pulse due to negative group velocity dispersion (GVD) and reaches the rear of the bubble before the laser pulse exits the plasma plateau. This process is illustrated in Fig.~\ref{fig_eta_n0}(b), where the longer-wavelength pulse corresponding to the generated THz radiation (marked by the black dashed contour) is seen to encounter the rear of the bubble at $x_p \approx 750~\mu$m. As a result, part of this longer-wavelength radiation undergoes photon acceleration, causing its wavenumber to upshift to values exceeding $k_0/3$, as also evident from the comparison between Fig.~\ref{fig_wigner}(c2) and Fig.~\ref{fig_wigner}(c3). 

The effect of the initial laser intensity (i.e., laser energy for a fixed spot size) on the laser-to-THz conversion efficiency has also been investigated. The corresponding PIC simulation results are presented in Fig.~\ref{fig_eta_a0} for a fixed initial plasma density of $n_0 = 0.05n_c$. It is observed that the net energy converted into THz radiation increases with increasing $a_0$, i.e., incident laser energy. However, the conversion efficiency does not increase proportionally. It is found to be lower for $a_0 = 7.5$ than for $a_0 = 5.0$, as shown in the inset of Fig.~\ref{fig_eta_a0}. This may be attributed to the fact that, with increasing $a_0$, other nonlinear processes, such as wave breaking and self-injection-induced electron acceleration, become increasingly dominant channels for laser energy depletion. 

It can be seen from Fig.~\ref{fig_eta_n0}(a) and Fig.~\ref{fig_eta_a0} that the conversion of laser energy into THz radiation in the frequency range $\nu < \nu_0/3 \approx 9.5$ THz begins only after the laser has propagated a certain distance through the plasma. The PIC simulations further show that the threshold propagation distance depends on the initial plasma density $n_0$, while remaining nearly independent of the laser intensity $a_0$. In addition, the laser-to-THz energy conversion efficiency saturates in some cases before the laser pulse reaches the end of the plasma plateau. These characteristic length scales observed in the simulations can be approximately estimated as follows. Assuming a plasma wake wave of the form $\delta n=\delta n_0 \sin(k_{p0}\zeta)$, where $k_{p0}=\omega_{p0}/c$, Eq.~\ref{eq1} can be written as

\begin{equation}
    \Delta \omega = -v_g\tau \frac{\omega_{p0}^2}{2\omega_0} k_{p0}\frac{\delta n_0}{n_0} \cos{(k_{p0}\zeta)}~~.
\label{eq6}    
\end{equation}

\noindent
In the bubble regime, where the laser ponderomotive force expels nearly all plasma electrons from the propagation axis, $\delta n_0 \sim n_0$. Under this assumption, the minimum propagation distance, $L_p=(v_g\tau)_m$, required for the laser frequency to be downshifted to $\omega_0/3$ can be expressed as

\begin{equation}
    L_p = \frac{4}{3}\frac{\omega_{0}^2}{\omega_{p0}^2} \frac{1}{k_{p0}}~~.
\label{eq7}    
\end{equation}

Once the laser frequency is downshifted, the generated low-frequency radiation begins to lag behind the main pulse due to the negative GVD of electromagnetic waves in plasma. Consequently, after propagating a certain distance, the generated THz pulse reaches the rear of the bubble, where a portion of it undergoes photon acceleration, leading to a decrease (or saturation) of the laser-to-THz energy conversion efficiency ($\eta$), as discussed above. The propagation distance $L_s$ at which this saturation begins can be estimated by considering a THz pulse with frequency $\nu=\nu_0/3$ generated near the front of the bubble. Saturation in $\eta$ is expected when this pulse slips backward by approximately one bubble length $\lambda_w \approx 2\pi c\sqrt{a_0}/\omega_{p}$ \cite{lu2007generating} and reaches the rear of the wake. Here, $\omega_p = \omega_{p0}/\sqrt{\gamma}$ is the relativistic mass-corrected electron plasma frequency. The corresponding distance can therefore be estimated as

\begin{equation}
    L_s = L_p + v_g\frac{\lambda_w}{v_g - v_g^{\rm THz}}~~,
\label{eq8}    
\end{equation}

\noindent
where $v_g \approx c\sqrt{1-\omega_{p0}^2/\omega_0^2}$ is the group velocity of the laser pulse in the plasma, and $v_g^{\rm THz} \approx c\sqrt{1-\omega_{p0}^2/\omega_{\rm THz}^2}$ is the corresponding group velocity of the generated THz radiation with frequency $\omega_{\rm THz}=\omega_0/3$.

Figure \ref{fig_theory} shows the variation of $L_p$ and $L_s$, calculated using Eqs. \ref{eq7} and \ref{eq8}, for different values of initial plasma density and $a_0$. To compare with the PIC simulations, an additional $200~\mu$m is added to both $L_p$ and $L_s$, since the plasma plateau in the simulations starts at $x_p = 200~\mu$m. As shown in Fig. \ref{fig_eta_n0}(a), $\eta$ becomes finite only after $x_p \simeq 300~\mu$m, $\simeq 400~\mu$m, and $\simeq 750~\mu$m for $n_0 = 0.08n_c$, $0.05n_c$, and $0.03n_c$, respectively. These values agree well with the corresponding values of $L_p$ shown in Fig. \ref{fig_theory}(a). Also, among these cases, only for $n_0 = 0.08n_c$, $L_s$ is smaller than the end of the plasma plateau ($x_p = 1025~\mu$m). This is consistent with the PIC results in Fig. \ref{fig_eta_n0}(a), where saturation of $\eta$ before the laser reaches the end of the plasma plateau is observed only for this case. For different values of $a_0$ at fixed $n_0 = 0.05n_c$, $L_p$ remains almost unchanged, while $L_s$ increases with $a_0$. It is seen that $L_s$ is smaller than the end location of the plasma plateau only for $a_0 = 2.5$. This is consistent with the PIC results in Fig. \ref{fig_eta_a0}, where THz generation starts at $x_p \simeq 400~\mu$m for all values of $a_0$. Moreover, saturation of $\eta$ is observed only for $a_0 = 2.5$, at $x_p \simeq 850~\mu$m, i.e., before the end of the plasma plateau.

In the present study, PIC simulations reveal a maximum energy conversion efficiency of the transmitted THz radiation of approximately $\eta \approx 4.0\%$. Assuming an input laser energy of $1.3$ J (corresponding to $a_0 = 5.0$ and $w_{0} = 30~\mu$m), the generated THz radiation may therefore contain an energy of approximately $50$ mJ. Furthermore, the conversion efficiency, and hence the generated THz energy, can be increased further by using a higher initial plasma density and an optimal plasma length, as discussed above.


\begin{figure}
  \centering
  \includegraphics[width=3.0in]{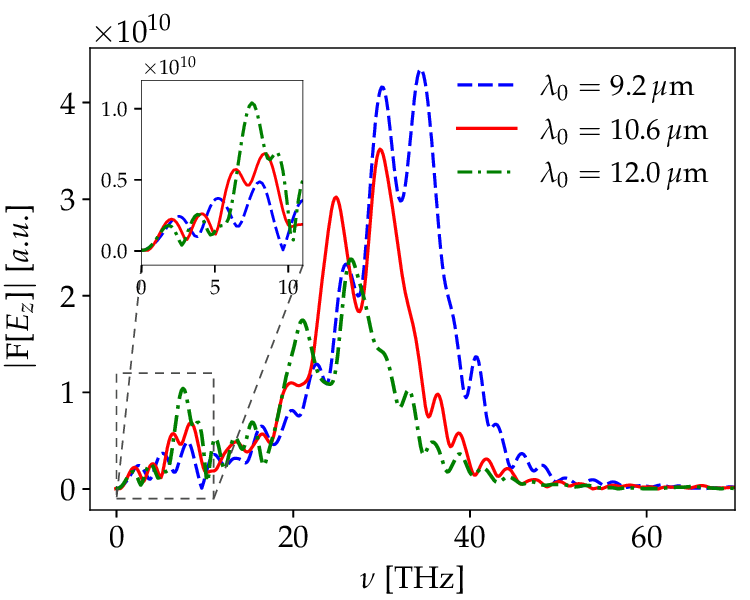}
  \caption{Fourier spectra of the on-axis $E_z(t)$ at a fixed position $x = 1200~\mu$m, i.e., $100~\mu$m beyond the plasma channel, for three different initial laser wavelengths $\lambda_0$. In all simulations, the peak value of the initial normalized vector potential ($a_0$) and the initial plasma density ($n_0$) were kept identical, with values of $a_0 = 5.0$ and $n_0 = 0.05n_c$, respectively.}
\label{fig_Thz_lmda}
\end{figure}

To examine the robustness of the photon-deceleration-driven THz generation mechanism for different initial laser wavelengths ($\lambda_0$) in the infrared regime, additional PIC simulations were performed by varying $\lambda_0$. In these simulations, all other parameters were kept identical, with fixed values of $a_0 = 5.0$ and $n_0 = 0.05n_c$. The results are shown in Fig. \ref{fig_Thz_lmda}. The Fourier spectra of the on-axis, $z$-polarized electric field $E_z(t)$ were obtained at a fixed position, $x = 1200~\mu$m, corresponding to $\sim 100~\mu$m beyond the plasma channel. A low-frequency ($<10$ THz) component is present for all cases, showing that the THz generation process remains effective over the considered range of $\lambda_0$. For fixed values of $a_0$, $w_0$, and $\tau_{\mathrm{fwhm}}$, the initial laser energy is higher for the shorter-wavelength cases. Despite this, the spectral power in the THz range is higher for the longer-wavelength case, with the $\lambda_0 = 12.0~\mu$m case showing the highest THz spectral power.

This dependence can be explained by the photon deceleration process. As seen from the Wigner spectrum in Fig. \ref{fig_wigner}, the laser frequency, and its wavenumber, continuously shifts to lower values as the laser propagates through the plasma. This downshift continues as long as the laser has sufficient energy to drive electron-density accumulation and maintain the refractive-index gradient. As the laser energy decreases, the density accumulation and refractive-index gradient become weaker, limiting further photon deceleration. Consequently, for a fixed propagation distance shorter than the pump depletion length, the initial laser wavelength influences the amount of spectral energy that can be transferred into the low-frequency THz range. A longer initial wavelength corresponds to a lower initial frequency and therefore requires a smaller frequency shift to reach the THz regime. As a result, the $\lambda_0 = 12.0~\mu$m case exhibits a higher THz spectral power than the shorter-wavelength cases, as shown in Fig. \ref{fig_Thz_lmda}.


\subsection{Three-Dimensional Effects}
\label{3D_section}


\begin{figure}
  \centering
  \includegraphics[width=3.0in]{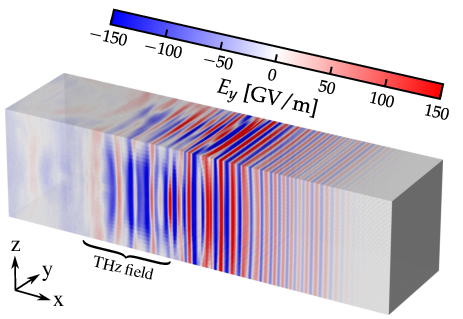}
  \caption{Electric field $E_y$ in 3D ($x$-$y$-$z$) space at a simulation time of $t = 4000$ fs, corresponding to a propagation distance of $\sim 1000~\mu$m, i.e., the end of the plasma plateau. Only a small cross-section of the transverse plane around the laser propagation axis and a short longitudinal section around the laser pulse are shown. The initial laser polarization is along $\hat{y}$. Longer-wavelength THz pulses with the same polarization are observed behind the laser electric field.}
\label{fig_3d_dens}
\end{figure}


\begin{figure}
  \centering
  \includegraphics[width=3.0in]{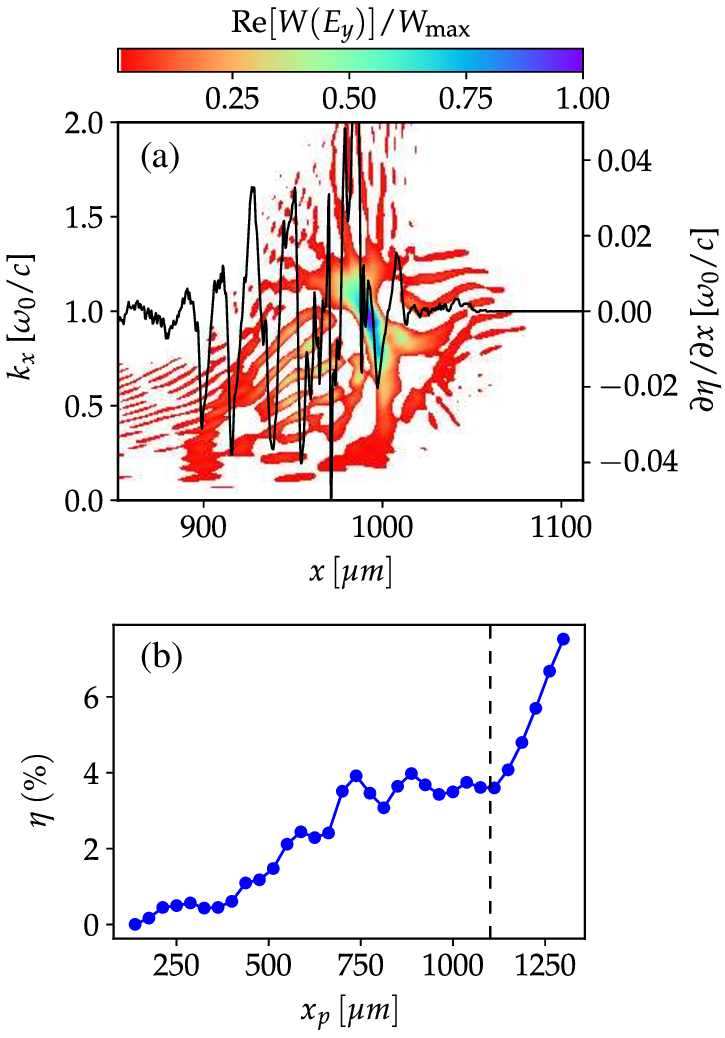}
  \caption{(a) Wigner spectrum (colorbar) of the on-axis electric field $E_y$, along with the spatial gradient of the refractive index calculated from the on-axis electron density at a simulation time of $t = 4000$ fs. (b) Laser-to-THz energy conversion efficiency, $\eta = \int |E_y^{\rm THz}|^2 dxdydz / \int |E_y^{\rm laser}|^2 dxdydz$, as a function of the laser propagation distance, $x_p = ct$. The vertical dashed line marks the plasma-vacuum boundary at the exit.}
\label{fig_3d_eta}
\end{figure}

To examine whether the photon-deceleration-driven THz generation mechanism is preserved in a fully three-dimensional (3D) geometry, an additional 3D PIC simulation was performed for a selected case. In this simulation, the peak normalized vector potential and initial plasma density are set to $a_0 = 5.0$ and $n_0 = 0.08n_c$, respectively. The laser and plasma profiles, as well as the simulation geometry, are taken to be identical in the two transverse directions. The initial laser electric field is polarized along $\hat y$, in contrast to the polarization used in the 2D simulation, to further demonstrate the polarization independence of the photon-deceleration mechanism. All other simulation parameters are kept identical to those of the corresponding 2D case, as described in Sec. \ref{simu}. The results of the 3D PIC simulation, shown in Figs. \ref{fig_3d_dens} and \ref{fig_3d_eta}, demonstrate the same main features of photon deceleration and THz generation observed in the 2D simulations.

The $y$-polarized electric field $E_y$ at $t = 4000$ fs, corresponding to the laser propagation up to the end of the plasma plateau, is shown in the 3D ($x$-$y$-$z$) plane in Fig. \ref{fig_3d_dens}. Longer-wavelength electric field pulses with the same polarization are observed behind the main laser pulse. These pulses correspond to the generated THz radiation, which lags behind the main laser pulse due to the negative GVD in plasma, as discussed in detail for the 2D case in Fig. \ref{fig_eta_n0}(b).

The photon deceleration process observed in the 3D PIC simulation is shown in Fig. \ref{fig_3d_eta}(a) through the Wigner spectrum of the on-axis $y$-polarized electric field $E_y$ at $t = 4000$ fs. The main features of the Wigner spectrum remain similar to those of the corresponding 2D case shown in Fig. \ref{fig_wigner}(c3). In particular, the wavenumber of the laser field is downshifted during propagation, and a part of the downshifted spectrum becomes separated from the main laser pulse. The laser-to-THz energy conversion efficiency ($\eta$), shown in Fig. \ref{fig_3d_eta}(b), also exhibits a similar trend to that observed in the 2D case. However, the conversion efficiency in the 3D case is slightly lower than that in the corresponding 2D case shown in Fig. \ref{fig_eta_n0}(a).

It is also observed that $\eta$ becomes finite from the beginning of the plasma target in the 3D case, unlike in the 2D case. This difference arises because the 3D calculation of $\eta$ also includes the contribution from the low-frequency wakefield, with a frequency around the plasma frequency, which starts to develop as soon as the laser enters the plasma. This contribution is absent in the 2D calculation because the out-of-plane electric field $E_z$ was used to calculate $\eta$. After the plasma-vacuum boundary, marked by the black dashed vertical line in Fig. \ref{fig_3d_eta}(b), $\eta$ increases again significantly. This increase is attributed to the contribution from the radially polarized THz radiation generated through coherent transition radiation (CTR) when the LWFA-accelerated electrons cross the plasma-vacuum boundary \cite{leemans2003observation, dechard2019thz}.

The lower conversion efficiency in the 3D case is mainly attributed to stronger laser divergence and weaker electron-density accumulation at the laser front. In 3D, electrons have an additional transverse degree of freedom, which reduces the electron-density accumulation at the laser front compared with the 2D case. This results in a weaker refractive-index gradient and, consequently, weaker photon deceleration, as seen in Fig. \ref{fig_3d_eta}(a) compared with Fig. \ref{fig_wigner}(c3). However, the conversion efficiency can be further enhanced by increasing the laser intensity ($a_0$) and/or the plasma density with an optimized plasma plateau length, as discussed in Sec. \ref{THz_section}.

\section{Summary and conclusions}
\label{summary}

The generation of high-energy, tunable terahertz (THz) radiation at frequencies below 10 THz has been demonstrated using long-wavelength infrared (LWIR) laser pulses interacting with a gaseous medium. Particle-in-cell (PIC) simulations reveal several important nonlinear effects during the interaction of the LWIR laser pulse with the gas target. In particular, in the self-modulated wakefield (SMWF) regime, where the plasma wavelength becomes shorter than the laser pulse length, the laser undergoes nonlinear modulation driven by the self-modulation instability. Interestingly, in addition to the Raman sidebands (Stokes and anti-Stokes), a longer-wavelength few-cycle pulse in the THz regime is generated and transmitted out of the medium behind the main laser pulse. The polarization of the transmitted THz pulse remains the same as that of the incident laser pulse.

Particle-in-cell (PIC) simulation results, supported by theoretical analysis, demonstrate that THz radiation is generated through the photon deceleration of an LWIR laser pulse as it interacts with the enhanced density accumulation in the wakefield under the SMWF regime. A theoretical scaling for the optimized plasma length required for efficient THz generation is obtained and validated through extensive PIC simulations over a range of initial target densities and laser intensities. The generated THz radiation has a broad spectrum extending below $10$ THz, demonstrating that photon deceleration can produce low-frequency THz radiation. The THz spectral peak can be tuned by varying the initial plasma density and the plasma propagation length. The simulations show that the laser-to-THz energy conversion efficiency of the transmitted radiation into vacuum can reach up to $4.0\%$, yielding 50 mJ of THz energy for the laser parameters considered in this study. Furthermore, it is found that a higher initial plasma density, along with an optimized plasma length, can further enhance the conversion efficiency. The conversion efficiency also increases with increasing laser intensity. The photon-deceleration-driven THz generation presented in this study can be experimentally investigated using the techniques described in Refs. \cite{nie2018relativistic, nie2020photon}. The findings of this study will be useful for the development of tunable and intense few-cycle THz sources in the future.


\section*{Acknowledgments}

The author acknowledges support from the INSPIRE Faculty Fellowship (Faculty Registration No. IFA24-PH 329) of the Department of Science $\&$ Technology, Government of India. The author thanks Dr. Garima Arora and Prof. Amita Das for the useful discussion. The computations were performed on the Param Vikram-1000 High Performance Computing Cluster of the Physical Research Laboratory (PRL).


\bibliography{ref}

@article{tonouchi2007cutting,
  title={Cutting-edge terahertz technology},
  author={Tonouchi, Masayoshi},
  journal={Nature photonics},
  volume={1},
  number={2},
  pages={97--105},
  year={2007},
  url = {https://doi.org/10.1038/nphoton.2007.3},
  doi = {10.1038/nphoton.2007.3},
  publisher={Nature Publishing Group UK London}
}

@article{baxter2011terahertz,
  title={Terahertz spectroscopy},
  author={Baxter, Jason B and Guglietta, Glenn W},
  journal={Analytical chemistry},
  volume={83},
  number={12},
  pages={4342--4368},
  year={2011},
  url = {https://doi.org/10.1021/ac200907z},
  doi = {10.1021/ac200907z},
  publisher={ACS Publications}
}

@article{kampfrath2013resonant,
  title={Resonant and nonresonant control over matter and light by intense terahertz transients},
  author={Kampfrath, Tobias and Tanaka, Koichiro and Nelson, Keith A},
  journal={Nature Photonics},
  volume={7},
  number={9},
  pages={680--690},
  year={2013},
  url = {https://doi.org/10.1038/nphoton.2013.184},
  doi = {10.1038/nphoton.2013.184},
  publisher={Nature Publishing Group UK London}
}

@article{berge2019terahertz,
  title={Terahertz spectroscopy from air plasmas created by two-color femtosecond laser pulses: The ALTESSE project},
  author={Berg{\'e}, L and Kaltenecker, K and Engelbrecht, S and Nguyen, A and Skupin, Stefan and Merlat, L and Fischer, B and Zhou, Binbin and Thiele, I and Jepsen, Peter Uhd},
  journal={Europhysics Letters},
  volume={126},
  number={2},
  pages={24001},
  year={2019},
  url = {https://doi.org/10.1209/0295-5075/126/24001},
  doi = {10.1209/0295-5075/126/24001},
  publisher={EDP Sciences, IOP Publishing and Societ{\`a} Italiana di Fisica}
}

@article{chen2007absorption,
  title={Absorption coefficients of selected explosives and related compounds in the range of 0.1--2.8 THz},
  author={Chen, Jian and Chen, Yunqing and Zhao, Hongwei and Bastiaans, Glenn J and Zhang, X-C},
  journal={Optics express},
  volume={15},
  number={19},
  pages={12060--12067},
  year={2007},
  url = {https://doi.org/10.1364/OE.15.012060},
  doi = {10.1364/OE.15.012060},
  publisher={Optical Society of America}
}

@article{davies2008terahertz,
  title={Terahertz spectroscopy of explosives and drugs},
  author={Davies, A Giles and Burnett, Andrew D and Fan, Wenhui and Linfield, Edmund H and Cunningham, John E},
  journal={Materials today},
  volume={11},
  number={3},
  pages={18--26},
  year={2008},
  url = {https://doi.org/10.1016/S1369-7021(08)70016-6},
  doi = {10.1016/S1369-7021(08)70016-6},
  publisher={Elsevier}
}

@inproceedings{kemp2003security,
  title={Security applications of terahertz technology},
  author={Kemp, Michael C and Taday, PF and Cole, Bryan E and Cluff, JA and Fitzgerald, Anthony J and Tribe, William R},
  booktitle={Terahertz for military and security applications},
  volume={5070},
  pages={44--52},
  year={2003},
  url = {https://doi.org/10.1117/12.500491},
  doi = {10.1117/12.500491},
  organization={SPIE}
}

@article{chan2007imaging,
  title={Imaging with terahertz radiation},
  author={Chan, Wai Lam and Deibel, Jason and Mittleman, Daniel M},
  journal={Reports on progress in physics},
  volume={70},
  number={8},
  pages={1325--1379},
  year={2007},
  url = {http://dx.doi.org/10.1088/0034-4885/70/8/R02},
  doi = {10.1088/0034-4885/70/8/R02},
  publisher={IOP Publishing}
}

@article{son2019potential,
  title={Potential clinical applications of terahertz radiation},
  author={Son, Joo-Hiuk and Oh, Seung Jae and Cheon, Hwayeong},
  journal={Journal of Applied Physics},
  volume={125},
  number={19},
  year={2019},
  url = {https://doi.org/10.1063/1.5080205},
  doi = {10.1063/1.5080205},
  publisher={AIP Publishing}
}

@article{clough2014toward,
  title={Toward remote sensing with broadband terahertz waves},
  author={Clough, Benjamin and Zhang, Xi-Cheng},
  journal={Frontiers of Optoelectronics},
  volume={7},
  number={2},
  pages={199--219},
  year={2014},
  url = {https://doi.org/10.1007/s12200-014-0397-3},
  doi = {10.1007/s12200-014-0397-3},
  publisher={Springer}
}

@article{hafez2016intense,
  title={Intense terahertz radiation and their applications},
  author={Hafez, HA and Chai, X and Ibrahim, A and Mondal, S and F{\'e}rachou, D and Ropagnol, X and Ozaki, T},
  journal={Journal of Optics},
  volume={18},
  number={9},
  pages={093004},
  year={2016},
  url = {http://dx.doi.org/10.1088/2040-8978/18/9/093004},
  doi = {10.1088/2040-8978/18/9/093004},
  publisher={IOP Publishing}
}

@article{crowe1992gaas,
  title={GaAs Schottky diodes for THz mixing applications},
  author={Crowe, Thomas W and Mattauch, Robert J and Roser, Hans Peter and Bishop, William L and Peatman, William CB and Liu, Xiaolei},
  journal={Proceedings of the IEEE},
  volume={80},
  number={11},
  pages={1827--1841},
  year={1992},
  url = {https://doi.org/10.1109/5.175258},
  doi = {10.1109/5.175258},
  publisher={IEEE}
}

@article{faist1994quantum,
  title={Quantum cascade laser},
  author={Faist, Jerome and Capasso, Federico and Sivco, Deborah L and Sirtori, Carlo and Hutchinson, Albert L and Cho, Alfred Y},
  journal={Science},
  volume={264},
  number={5158},
  pages={553--556},
  year={1994},
  url = {https://doi.org/10.1126/science.264.5158.553},
  doi = {10.1126/science.264.5158.553},
  publisher={American Association for the Advancement of Science}
}

@article{loffler2005large,
  title={Large-area electro-optic ZnTe terahertz emitters},
  author={L{\"o}ffler, T and Hahn, T and Thomson, M and Jacob, F and Roskos, HG},
  journal={Optics express},
  volume={13},
  number={14},
  pages={5353--5362},
  year={2005},
  url = {https://doi.org/10.1364/OPEX.13.005353},
  doi = {10.1364/OPEX.13.005353},
  publisher={Optical Society of America}
}

@article{hebling2004tunable,
  title={Tunable THz pulse generation by optical rectification of ultrashort laser pulses with tilted pulse fronts},
  author={Hebling, J and Stepanov, AG and Alm{\'a}si, G and Bartal, B and Kuhl, J},
  journal={Applied Physics B},
  volume={78},
  number={5},
  pages={593--599},
  year={2004},
  url = {https://doi.org/10.1007/s00340-004-1469-7},
  doi = {10.1007/s00340-004-1469-7},
  publisher={Springer}
}

@article{vicario2014gv,
  title={GV/m single-cycle terahertz fields from a laser-driven large-size partitioned organic crystal},
  author={Vicario, Carlo and Monoszlai, Balazs and Hauri, Christoph P},
  journal={Physical review letters},
  volume={112},
  number={21},
  pages={213901},
  year={2014},
  url = {https://doi.org/10.1103/PhysRevLett.112.213901},
  doi = {10.1103/PhysRevLett.112.213901},
  publisher={APS}
}

@article{hamster1993subpicosecond,
  title={Subpicosecond, electromagnetic pulses from intense laser-plasma interaction},
  author={Hamster, H and Sullivan, A and Gordon, S and White, W and Falcone, RW},
  journal={Physical review letters},
  volume={71},
  number={17},
  pages={2725},
  year={1993},
  url = {https://doi.org/10.1103/PhysRevLett.71.2725},
  doi = {10.1103/PhysRevLett.71.2725},
  publisher={APS}
}

@article{herzer2018investigation,
  title={An investigation on THz yield from laser-produced solid density plasmas at relativistic laser intensities},
  author={Herzer, S and Woldegeorgis, A and Polz, J and Reinhard, A and Almassarani, M and Beleites, B and Ronneberger, F and Grosse, R and Paulus, GG and H{\"u}bner, U and others},
  journal={New Journal of Physics},
  volume={20},
  number={6},
  pages={063019},
  year={2018},
  url = {https://doi.org/10.1088/1367-2630/aaada0},
  doi = {10.1088/1367-2630/aaada0},
  publisher={IOP Publishing}
}

@article{liao2015bursts,
  title={Bursts of terahertz radiation from large-scale plasmas irradiated by relativistic picosecond laser pulses},
  author={Liao, GQ and Li, YT and Li, C and Su, LN and Zheng, Y and Liu, M and Wang, WM and Hu, ZD and Yan, WC and Dunn, J and others},
  journal={Physical Review Letters},
  volume={114},
  number={25},
  pages={255001},
  year={2015},
  url = {https://doi.org/10.1103/PhysRevLett.114.255001},
  doi = {10.1103/PhysRevLett.114.255001},
  publisher={APS}
}

@article{gopal2013observation,
  title={Observation of gigawatt-class THz pulses from a compact laser-driven particle accelerator},
  author={Gopal, A and Herzer, S and Schmidt, A and Singh, P and Reinhard, A and Ziegler, W and Br{\"o}mmel, D and Karmakar, A and Gibbon, Paul and Dillner, U and others},
  journal={Physical Review Letters},
  volume={111},
  number={7},
  pages={074802},
  year={2013},
  url = {http://dx.doi.org/10.1103/PhysRevLett.111.074802},
  doi = {10.1103/PhysRevLett.111.074802},
  publisher={APS}
}

@article{cook2000intense,
  title={Intense terahertz pulses by four-wave rectification in air},
  author={Cook, DJ and Hochstrasser, RM},
  journal={Optics letters},
  volume={25},
  number={16},
  pages={1210--1212},
  year={2000},
  url = {https://doi.org/10.1364/OL.25.001210},
  doi = {10.1364/OL.25.001210},
  publisher={Optical Society of America}
}

@article{kim2008coherent,
  title={Coherent control of terahertz supercontinuum generation in ultrafast laser--gas interactions},
  author={Kim, Ki-Yong and Taylor, Antoinette Jane and Glownia, JH and Rodriguez, George},
  journal={Nature photonics},
  volume={2},
  number={10},
  pages={605--609},
  year={2008},
  url = {http://www.nature.com/doifinder/10.1038/nphoton.2008.153},
  doi = {10.1038/nphoton.2008.153},
  publisher={Nature Publishing Group}
}

@article{meng2016enhancement,
  title={Enhancement of terahertz radiation by using circularly polarized two-color laser fields},
  author={Meng, Chao and Chen, Wenbo and Wang, Xiaowei and L{\"u}, Zhihui and Huang, Yindong and Liu, Jinlei and Zhang, Dongwen and Zhao, Zengxiu and Yuan, Jianmin},
  journal={Applied Physics Letters},
  volume={109},
  number={13},
  year={2016},
  url = {https://doi.org/10.1063/1.4963883},
  doi = {10.1063/1.4963883},
  publisher={AIP Publishing}
}

@article{clerici2013wavelength,
  title={Wavelength scaling of terahertz generation by gas ionization},
  author={Clerici, Matteo and Peccianti, Marco and Schmidt, Bruno E and Caspani, Lucia and Shalaby, Mostafa and Gigu{\`e}re, Mathieu and Lotti, Antonio and Couairon, Arnaud and L{\'e}gar{\'e}, Fran{\c{c}}ois and Ozaki, Tsuneyuki and others},
  journal={Physical Review Letters},
  volume={110},
  number={25},
  pages={253901},
  year={2013},
  url = {https://doi.org/10.1103/PhysRevLett.110.253901},
  doi = {10.1103/PhysRevLett.110.253901},
  publisher={APS}
}

@article{chen2015high,
  title={High field terahertz emission from relativistic laser-driven plasma wakefields},
  author={Chen, Zi-Yu and Pukhov, Alexander},
  journal={Physics of Plasmas},
  volume={22},
  number={10},
  year={2015},
  url = {https://doi.org/10.1063/1.4933130},
  doi = {10.1063/1.4933130},
  publisher={AIP Publishing}
}

@article{Tajima1979,
  title = {Laser Electron Accelerator},
  author = {Tajima, T. and Dawson, J. M.},
  journal = {Phys. Rev. Lett.},
  volume = {43},
  issue = {4},
  pages = {267--270},
  numpages = {0},
  year = {1979},
  month = {Jul},
  publisher = {American Physical Society},
  doi = {10.1103/PhysRevLett.43.267},
  url = {https://link.aps.org/doi/10.1103/PhysRevLett.43.267}
}

@article{RevModPhys.81.1229,
  title = {Physics of laser-driven plasma-based electron accelerators},
  author = {Esarey, E. and Schroeder, C. B. and Leemans, W. P.},
  journal = {Rev. Mod. Phys.},
  volume = {81},
  issue = {3},
  pages = {1229--1285},
  numpages = {0},
  year = {2009},
  month = {Aug},
  publisher = {American Physical Society},
  doi = {10.1103/RevModPhys.81.1229},
  url = {https://link.aps.org/doi/10.1103/RevModPhys.81.1229}
}

@article{maity2024parametric,
  title={Parametric analysis of electron beam quality in laser wakefield acceleration based on the truncated ionization injection mechanism},
  author={Maity, Srimanta and Mondal, Alamgir and Vishnyakov, Eugene and Molodozhentsev, Alexander},
  journal={Plasma Physics and Controlled Fusion},
  volume={66},
  number={3},
  pages={035012},
  year={2024},
  url = {https://doi.org/10.1088/1361-6587/ad238e},
  doi = {10.1088/1361-6587/ad238e},
  publisher={IOP Publishing}
}

@article{maity2025coupling,
  title={Coupling and acceleration of externally injected electron beams in laser-driven plasma wakefields},
  author={Maity, Srimanta and Sasorov, Pavel and Molodozhentsev, Alexander},
  journal={Journal of Physics D: Applied Physics},
  volume={58},
  number={14},
  pages={145204},
  year={2025},
  publisher={IOP Publishing},
  doi = {10.1088/1361-6463/adb6b9},
  url = {https://doi.org/10.1088/1361-6463/adb6b9}
}

@article{maity2026laser,
  title={Laser wakefield acceleration in a capillary gas cell to produce high-quality GeV-scale electron beams},
  author={Maity, Srimanta and Massimo, Francesco and Whitehead, Alex and Sasorov, Pavel and Molodozhentsev, Alexander},
  journal={Physical Review Accelerators and Beams},
  volume={29},
  number={9},
  pages={093601},
  year={2026},
  url = {https://doi.org/10.1103/l8b2-j67x},
  doi = {10.1103/l8b2-j67x},
  publisher={APS}
}

@article{leemans2003observation,
  title={Observation of terahertz emission from a laser-plasma accelerated electron bunch crossing a plasma-vacuum boundary},
  author={Leemans, WP and Geddes, CGR and Faure, J and T{\'o}th, Cs and Van Tilborg, J and Schroeder, CB and Esarey, E and Fubiani, G and Auerbach, D and Marcelis, B and others},
  journal={Physical review letters},
  volume={91},
  number={7},
  pages={074802},
  year={2003},
  url = {https://doi.org/10.1103/PhysRevLett.91.074802},
  doi = {10.1103/PhysRevLett.91.074802},
  publisher={APS}
}

@article{leemans2004terahertz,
  title={Terahertz radiation from laser accelerated electron bunches},
  author={Leemans, WP and Van Tilborg, J and Faure, J and Geddes, CGR and T{\'o}th, Cs and Schroeder, CB and Esarey, E and Fubiani, G and Dugan, G},
  journal={Physics of plasmas},
  volume={11},
  number={5},
  pages={2899--2906},
  year={2004},
  url = {https://doi.org/10.1063/1.1652834},
  doi = {10.1063/1.1652834},
  publisher={American Institute of Physics}
}

@article{dechard2018terahertz,
  title={Terahertz pulse generation in underdense relativistic plasmas: From photoionization-induced radiation to coherent transition radiation},
  author={D{\'e}chard, J and Debayle, A and Davoine, X and Gremillet, L and Berg{\'e}, L},
  journal={Physical review letters},
  volume={120},
  number={14},
  pages={144801},
  year={2018},
  url = {https://doi.org/10.1103/PhysRevLett.120.144801},
  doi = {10.1103/PhysRevLett.120.144801},
  publisher={APS}
}

@article{sheng2005emission,
  title={Emission of electromagnetic pulses from laser wakefields through linear mode conversion},
  author={Sheng, Zheng-Ming and Mima, Kunioki and Zhang, Jie and Sanuki, Heiji},
  journal={Physical review letters},
  volume={94},
  number={9},
  pages={095003},
  year={2005},
  url = {https://doi.org/10.1103/PhysRevLett.94.095003},
  doi = {10.1103/PhysRevLett.94.095003},
  publisher={APS}
}

@article{wang2024millijoule,
  title={Millijoule terahertz radiation from laser wakefields in nonuniform plasmas},
  author={Wang, Linzheng and Zhang, Zhelin and Chen, Siyu and Chen, Yanping and Hu, Xichen and Zhu, Mingyang and Yan, Wenchao and Xu, Hao and Sun, Lili and Chen, Min and others},
  journal={Physical Review Letters},
  volume={132},
  number={16},
  pages={165002},
  year={2024},
  url = {https://doi.org/10.1103/PhysRevLett.132.165002},
  doi = {10.1103/PhysRevLett.132.165002},
  publisher={APS}
}

@article{nguyen2018broadband,
  title={Broadband terahertz radiation from two-color mid-and far-infrared laser filaments in air},
  author={Nguyen, Alis{\'e}e and Gonz{\'a}lez de Alaiza Mart{\'\i}nez, Pedro and Thiele, Illia and Skupin, Stefan and Berg{\'e}, Luc},
  journal={Physical Review A},
  volume={97},
  number={6},
  pages={063839},
  year={2018},
  url = {https://doi.org/10.1103/PhysRevA.97.063839},
  doi = {10.1103/PhysRevA.97.063839},
  publisher={APS}
}

@article{PhysRevA.98.053415,
  title = {Boosting terahertz-radiation power with two-color circularly polarized midinfrared laser pulses},
  author = {Tulsky, V. A. and Baghery, M. and Saalmann, U. and Popruzhenko, S. V.},
  journal = {Phys. Rev. A},
  volume = {98},
  issue = {5},
  pages = {053415},
  year = {2018},
  url = {https://link.aps.org/doi/10.1103/PhysRevA.98.053415},
  doi = {10.1103/PhysRevA.98.053415},
  publisher = {American Physical Society}
}

@article{dechard2019thz,
  title={THz generation from relativistic plasmas driven by near-to far-infrared laser pulses},
  author={D{\'e}chard, J and Davoine, X and Berg{\'e}, L},
  journal={Physical Review Letters},
  volume={123},
  number={26},
  pages={264801},
  year={2019},
  url = {https://doi.org/10.1103/PhysRevLett.123.264801},
  doi = {10.1103/PhysRevLett.123.264801},
  publisher={APS}
}

@article{maity2025enhanced,
  title={Enhanced terahertz emission from the wakefield of CO 2-laser-created plasma},
  author={Maity, Srimanta and Arora, Garima},
  journal={Physical Review E},
  volume={111},
  number={4},
  pages={045205},
  year={2025},
  url = {https://doi.org/10.1103/PhysRevE.111.045205},
  doi = {10.1103/PhysRevE.111.045205},
  publisher={APS}
}

@article{esarey1990frequency,
  title={Frequency shifts induced in laser pulses by plasma waves},
  author={Esarey, E and Ting, A and Sprangle, P},
  journal={Physical Review A},
  volume={42},
  number={6},
  pages={3526},
  year={1990},
  url = {https://doi.org/10.1103/PhysRevA.42.3526},
  doi = {10.1103/PhysRevA.42.3526},
  publisher={APS}
}

@article{nie2018relativistic,
  title={Relativistic single-cycle tunable infrared pulses generated from a tailored plasma density structure},
  author={Nie, Zan and Pai, Chih-Hao and Hua, Jianfei and Zhang, Chaojie and Wu, Yipeng and Wan, Yang and Li, Fei and Zhang, Jie and Cheng, Zhi and Su, Qianqian and others},
  journal={Nature Photonics},
  volume={12},
  number={8},
  pages={489--494},
  year={2018},
  url = {https://doi.org/10.1038/s41566-018-0190-8},
  doi = {10.1038/s41566-018-0190-8},
  publisher={Nature Publishing Group UK London}
}

@article{nie2020photon,
  title={Photon deceleration in plasma wakes generates single-cycle relativistic tunable infrared pulses},
  author={Nie, Zan and Pai, Chih-Hao and Zhang, Jie and Ning, Xiaonan and Hua, Jianfei and He, Yunxiao and Wu, Yipeng and Su, Qianqian and Liu, Shuang and Ma, Yue and others},
  journal={Nature Communications},
  volume={11},
  number={1},
  pages={2787},
  year={2020},
  url = {https://doi.org/10.1038/s41467-020-16541-w},
  doi = {10.1038/s41467-020-16541-w},
  publisher={Nature Publishing Group UK London}
}

@article{zhu2022generation,
  title={Generation of single-cycle relativistic infrared pulses at wavelengths above 20 $\mu$m from density-tailored plasmas},
  author={Zhu, Xing-Long and Liu, Wei-Yuan and Weng, Su-Ming and Chen, Min and Sheng, Zheng-Ming and Zhang, Jie},
  journal={Matter and Radiation at Extremes},
  volume={7},
  number={1},
  year={2022},
  url = {https://doi.org/10.1063/5.0068265},
  doi = {10.1063/5.0068265},
  publisher={AIP Publishing}
}

@article{polyanskiy2015chirped,
  title={Chirped-pulse amplification in a CO2 laser},
  author={Polyanskiy, Mikhail N and Babzien, Marcus and Pogorelsky, Igor V},
  journal={Optica},
  volume={2},
  number={8},
  pages={675--681},
  year={2015},
  url = {https://doi.org/10.1364/OPTICA.2.000675},
  doi = {10.1364/OPTICA.2.000675},
  publisher={Optical Society of America}
}

@article{polyanskiy2020demonstration,
  title={Demonstration of a 2 ps, 5 TW peak power, long-wave infrared laser based on chirped-pulse amplification with mixed-isotope CO 2 amplifiers},
  author={Polyanskiy, Mikhail N and Pogorelsky, Igor V and Babzien, Marcus and Palmer, Mark A},
  journal={OSA Continuum},
  volume={3},
  number={3},
  pages={459--472},
  year={2020},
  doi = {10.1364/OSAC.381467},
  url = {https://doi.org/10.1364/OSAC.381467},
  publisher={Optica Publishing Group}
}

@article{panagiotopoulos2020multi,
  title={Multi-terawatt femtosecond 10 $\mu$m laser pulses by self-compression in a CO2 cell},
  author={Panagiotopoulos, Paris and Hastings, Michael G and Kolesik, Miroslav and Tochitsky, Sergei and Moloney, Jerome V},
  journal={OSA Continuum},
  volume={3},
  number={11},
  pages={3040--3047},
  year={2020},
  url = {https://doi.org/10.1364/OSAC.399992},
  doi = {10.1364/OSAC.399992},
  publisher={Optical Society of America}
}

@article{arber2015contemporary,
  title={Contemporary particle-in-cell approach to laser-plasma modelling},
  author={Arber, T. D. and Bennett, K. and Brady, C. S. and Lawrence-Douglas, A. and Ramsay, M. G. and Sircombe, N. J. and Gillies, P. and Evans, R. G. and Schmitz, H. and Bell, A. R. and others},
  journal={Plasma Physics and Controlled Fusion},
  volume={57},
  number={11},
  pages={113001},
  year={2015},
  url = {http://dx.doi.org/10.1088/0741-3335/57/11/113001},
  doi = {10.1088/0741-3335/57/11/113001},
  publisher={IOP Publishing}
}

@article{bennett2017users,
  title={Users Manual for the EPOCH PIC codes},
  author={Bennett, K. and Brady, C. and Schmitz, H. and Ridgers, C. and Arber, T. and Evans, R. and Bell, T.},
  journal={University of Warwick},
  url = {https://www.archie-west.ac.uk/wp-content/uploads/2014/02/epoch_user-4.3.pdf},
  year={2017}
}

@article{yee1966numerical,
  title={Numerical solution of initial boundary value problems involving Maxwell's equations in isotropic media},
  author={Yee, Kane},
  journal={IEEE Transactions on antennas and propagation},
  volume={14},
  number={3},
  pages={302--307},
  year={1966},
  url = {http://dx.doi.org/10.1109/TAP.1966.1138693},
  doi={10.1109/TAP.1966.1138693},
  publisher={IEEE}
}

@inproceedings{boris1970relativistic,
  title={Relativistic plasma simulation-optimization of a hybrid code},
  author={Boris, Jay P and others},
  booktitle={Proc. 4th Conf. on Numerical Simulation of
Plasmas (Washington, DC)},
  pages={3--67},
  year={1970}
}

@article{courant1967partial,
  title={On the partial difference equations of mathematical physics},
  author={Courant, Richard and Friedrichs, Kurt and Lewy, Hans},
  journal={IBM journal of Research and Development},
  volume={11},
  number={2},
  pages={215--234},
  year={1967},
  url = {https://web.stanford.edu/class/cme324/classics/courant-friedrichs-lewy.pdf},
  publisher={IBM}
}

@book{delone2000multiphoton,
  title={Multiphoton processes in atoms},
  author={Delone, Nikolai B and Krainov, Vladimir P},
  volume={13},
  year={2000},
  url = {https://doi.org/10.1007/978-3-642-57208-1},
  doi={10.1007/978-3-642-57208-1},
  publisher={Springer Berlin, Heidelberg}
}

@article{ammosov1986tunnel,
  title={Tunnel ionization of complex atoms and of atomic ions in an alternating electromagnetic field},
  author={Ammosov, Maxim V and Delone, Nikolai B and Krainov, Vladimir P},
  journal={Soviet Journal of Experimental and Theoretical Physics},
  volume={64},
  number={6},
  pages={1191},
  year={1986},
  url = {http://jetp.ras.ru/cgi-bin/dn/e_064_06_1191},
  publisher = {Maik Nauka/Interperiodica (Russia)}
}

@article{krainov1995theory,
  title={Theory of barrier-suppression ionization of atoms},
  author={Krainov, Vladimir P},
  journal={Journal of Nonlinear Optical Physics \& Materials},
  volume={4},
  number={04},
  pages={775--798},
  year={1995},
  url = {https://doi.org/10.1142/S0218863595000343},
  doi={10.1142/S0218863595000343},
  publisher={World Scientific}
}

@article{sprangle1990nonlinear,
  title={Nonlinear interaction of intense laser pulses in plasmas},
  author={Sprangle, P and Esarey, E and Ting, A},
  journal={Physical review A},
  volume={41},
  number={8},
  pages={4463},
  year={1990},
  url = {https://doi.org/10.1103/PhysRevA.41.4463},
  doi={10.1103/PhysRevA.41.4463},
  publisher={APS}
}

@article{bulanov1995two,
  title={Two-dimensional regimes of self-focusing, wake field generation, and induced focusing of a short intense laser pulse in an underdense plasma},
  author={Bulanov, SV and Pegoraro, Francesco and Pukhov, AM},
  journal={Physical review letters},
  volume={74},
  number={5},
  pages={710},
  year={1995},
  url = {https://doi.org/10.1103/PhysRevLett.74.710},
  doi={10.1103/PhysRevLett.74.710},
  publisher={APS}
}

@article{pukhov2002laser,
  title={Laser wake field acceleration: the highly non-linear broken-wave regime},
  author={Pukhov, Alexancer and Meyer-ter-Vehn, J{\"u}rgen},
  journal={Applied Physics B},
  volume={74},
  number={4},
  pages={355--361},
  year={2002},
  url = {https://doi.org/10.1007/s003400200795},
  doi={10.1007/s003400200795},
  publisher={Springer}
}

@article{joshi1981forward,
  title={Forward Raman instability and electron acceleration},
  author={Joshi, C and Tajima, T and Dawson, JM and Baldis, HA and Ebrahim, NA},
  journal={Physical Review Letters},
  volume={47},
  number={18},
  pages={1285},
  year={1981},
  url = {https://doi.org/10.1103/PhysRevLett.47.1285},
  doi={10.1103/PhysRevLett.47.1285},
  publisher={APS}
}

@article{andreev1992resonant,
  title={Resonant excitation of wakefields by a laser pulse in a plasma},
  author={Andreev, NE and Gorbunov, LM and Kirsanov, VI and Pogosova, AA and Ramazashvili, RR},
  journal={JETP lett},
  volume={55},
  number={10},
  pages={571--576},
  year={1992},
  url = {http://jetpletters.ru/ps/1277/article_19308.pdf}
}

@article{esarey1994envelope,
  title={Envelope analysis of intense laser pulse self-modulation in plasmas},
  author={Esarey, Eric and Krall, Jonathan and Sprangle, Phillip},
  journal={Physical review letters},
  volume={72},
  number={18},
  pages={2887},
  year={1994},
  url = {https://doi.org/10.1103/PhysRevLett.72.2887},
  doi={10.1103/PhysRevLett.72.2887},
  publisher={APS}
}

@article{zgadzaj2024plasma,
  title={Plasma electron acceleration driven by a long-wave-infrared laser},
  author={Zgadzaj, R and Welch, J and Cao, Y and Amorim, LD and Cheng, Aiqi and Gaikwad, Apurva and Iapozzutto, P and Kumar, Prabhat and Litvinenko, VN and Petrushina, I and others},
  journal={Nature Communications},
  volume={15},
  number={1},
  pages={4037},
  year={2024},
  url = {https://doi.org/10.1038/s41467-024-48413-y},
  doi={10.1038/s41467-024-48413-y},
  publisher={Nature Publishing Group UK London}
}

@article{antonsen1992self,
  title={Self-focusing and Raman scattering of laser pulses in tenuous plasmas},
  author={Antonsen Jr, TM and Mora, P},
  journal={Physical review letters},
  volume={69},
  number={15},
  pages={2204},
  year={1992},
  url = {https://doi.org/10.1103/PhysRevLett.69.2204},
  doi={10.1103/PhysRevLett.69.2204},
  publisher={APS}
}

@article{sprangle1992propagation,
  title={Propagation and guiding of intense laser pulses in plasmas},
  author={Sprangle, Phillip and Esarey, Eric and Krall, Jonathan and Joyce, G},
  journal={Physical review letters},
  volume={69},
  number={15},
  pages={2200},
  year={1992},
  url = {https://doi.org/10.1103/PhysRevLett.69.2200},
  doi= {10.1103/PhysRevLett.69.2200},
  publisher={APS}
}

@article{lu2007generating,
  title={Generating multi-GeV electron bunches using single stage laser wakefield acceleration in a 3D nonlinear regime},
  author={Lu, Wei and Tzoufras, M and Joshi, C and Tsung, FS and Mori, WB and Vieira, J and Fonseca, RA and Silva, LO},
  journal={Physical Review Special Topics—Accelerators and Beams},
  volume={10},
  number={6},
  pages={061301},
  year={2007},
  url = {https://doi.org/10.1103/PhysRevSTAB.10.061301},
  doi={10.1103/PhysRevSTAB.10.061301},
  publisher={APS}
}

@article{fulop2020laser,
  title={Laser-driven strong-field terahertz sources},
  author={F{\"u}l{\"o}p, J{\'o}zsef Andr{\'a}s and Tzortzakis, Stelios and Kampfrath, Tobias},
  journal={Advanced Optical Materials},
  volume={8},
  number={3},
  pages={1900681},
  year={2020},
  url = {https://doi.org/10.1002/adom.201900681},
  doi={10.1002/adom.201900681},
  publisher={Wiley Online Library}
}

@article{liao2023perspectives,
  title={Perspectives on ultraintense laser-driven terahertz radiation from plasmas},
  author={Liao, Guoqian and Li, Yutong},
  journal={Physics of Plasmas},
  volume={30},
  number={9},
  year={2023},
  url = {https://doi.org/10.1063/5.0167730},
  doi={10.1063/5.0167730},
  publisher={AIP Publishing}
}

@article{lemery2020highly,
  title={Highly scalable multicycle THz production with a homemade periodically poled macrocrystal},
  author={Lemery, Fran{\c{c}}ois and Vinatier, Thomas and Mayet, Frank and A{\ss}mann, Ralph and Baynard, Elsa and Demailly, Julien and Dorda, Ulrich and Lucas, Bruno and Pandey, Alok-Kumar and Pittman, Moana},
  journal={Communications Physics},
  volume={3},
  number={1},
  pages={150},
  year={2020},
  url = {https://doi.org/10.1038/s42005-020-00421-2},
  doi={10.1038/s42005-020-00421-2},
  publisher={Nature Publishing Group UK London}
}

@article{wu2023generation,
  title={Generation of 13.9-mJ terahertz radiation from lithium niobate materials},
  author={Wu, Xiaojun and Kong, Deyin and Hao, Sibo and Zeng, Yushan and Yu, Xieqiu and Zhang, Baolong and Dai, Mingcong and Liu, Shaojie and Wang, Jiaqi and Ren, Zejun and others},
  journal={Advanced Materials},
  volume={35},
  number={23},
  pages={2208947},
  year={2023},
  url = {https://doi.org/10.1002/adma.202208947},
  doi={10.1002/adma.202208947},
  publisher={Wiley Online Library}
}

@article{liao2020towards,
  title={Towards terawatt-scale spectrally tunable terahertz pulses via relativistic laser-foil interactions},
  author={Liao, Guo-Qian and Liu, Hao and Scott, Graeme G and Zhang, Yi-Hang and Zhu, Bao-Jun and Zhang, Zhe and Li, Yu-Tong and Armstrong, Chris and Zemaityte, Egle and Bradford, Philip and others},
  journal={Physical Review X},
  volume={10},
  number={3},
  pages={031062},
  year={2020},
  url = {https://doi.org/10.1103/PhysRevX.10.031062},
  doi={10.1103/PhysRevX.10.031062},
  publisher={APS}
}

@article{ruan2026terawatt,
  title={Terawatt-level widely-tunable terahertz pulses from femtosecond laser-irradiated metallic foils},
  author={Ruan, Jieya and Zhang, Xinyao and Liao, Guoqian and Sun, Fangzheng and Wei, Yanyu and Chen, Hao and Dong, Yichen and Qiu, Zhijie and Wang, Jinguang and Li, Yifei and others},
  journal={Optica},
  volume={13},
  number={4},
  pages={669--676},
  year={2026},
  url = {https://doi.org/10.1364/OPTICA.581695},
  doi={10.1364/OPTICA.581695},
  publisher={Optica Publishing Group}
}

@article{peng2023amplification,
  title={Amplification of a terahertz wave via stimulated Raman scattering},
  author={Peng, N and Chen, Y and Chen, YG and Tan, S and Yao, WB and Li, YX and Yu, JQ and Xiao, CZ},
  journal={Optics Letters},
  volume={48},
  number={9},
  pages={2433--2436},
  year={2023},
  url = {https://doi.org/10.1364/OL.484033},
  doi={10.1364/OL.484033},
  publisher={Optica Publishing Group}
}

@article{lu2025generation,
  title={Generation of multi-cycle relativistic terahertz radiation from photon deceleration in a long self-induced plasma wake},
  author={Lu, Huangang and Liu, Jiaxin and Cai, Jie and Wu, Xuezhi and Zhang, Jianyao and Ma, Qianyi and Wang, Zhenan and Xia, Yuhui and Ma, Wenjun and Lin, Chen and others},
  journal={Physics of Plasmas},
  volume={32},
  number={2},
  year={2025},
  publisher={AIP Publishing}
}

@article{pak2023multi,
  title={Multi-millijoule terahertz emission from laser-wakefield-accelerated electrons},
  author={Pak, Taegyu and Rezaei-Pandari, Mohammad and Kim, Sang Beom and Lee, Geonwoo and Wi, Dae Hee and Hojbota, Calin Ioan and Mirzaie, Mohammad and Kim, Hyeongmun and Sung, Jae Hee and Lee, Seong Ku and others},
  journal={Light: Science \& Applications},
  volume={12},
  number={1},
  pages={37},
  year={2023},
  url = {https://doi.org/10.1038/s41377-022-01068-0},
  doi = {10.1038/s41377-022-01068-0},
  publisher={Nature Publishing Group UK London}
}
\end{document}